\documentstyle[12pt,psfig]{article}
\input xy
\xyoption{all}

\newcommand\ext[1]{{\rm Ext}^{#1}}
\newcommand\uext{\underline{{\rm Ext}}}

\begin{document}

\hfill ILL-(TH)-02-06

\hfill hep-th/0208104

\vspace{0.75in}

\begin{center}

{\large\bf D-branes, open string vertex operators,}

{\large\bf and Ext groups}

\vspace{0.25in}

Sheldon Katz$^{1,2}$ and Eric Sharpe$^{1}$ \\
$^1$ Department of Mathematics \\
1409 W. Green St., MC-382 \\
University of Illinois \\
Urbana, IL  61801 \\
$^2$ Department of Physics\\
University of Illinois at Urbana-Champaign\\
Urbana, IL 61801\\
{\tt katz@math.uiuc.edu},
{\tt ersharpe@uiuc.edu} \\

 $\,$

\end{center}

In this note we explicitly work out the precise
relationship between Ext groups and massless modes
of D-branes wrapped on complex submanifolds of
Calabi-Yau manifolds.  
Specifically, we explicitly compute the boundary vertex operators
for massless Ramond sector states, in open string B models
describing Calabi-Yau manifolds at large radius,
directly in BCFT using standard methods.
Naively these vertex operators are in one-to-one
correspondence with certain sheaf cohomology
groups (as is typical for such
vertex operator calculations), which are related to the desired
Ext groups via spectral sequences.
However, a subtlety in the physics of the
open string B model has the effect
of physically realizing those spectral sequences
in BRST cohomology, so that the vertex operators
are actually in one-to-one correspondence
with Ext group elements.
This gives an extremely concrete physical test of recent
proposals regarding the relationship between
derived categories and D-branes.
The Freed-Witten anomaly also plays an important role in these
calculations, and we are now able to completely reconcile
that anomaly with the derived categories program generally.
We check these results extensively in numerous examples,
and comment on several related issues.

\begin{flushleft}
August 2002
\end{flushleft}

\newpage

\tableofcontents

\newpage

\section{Introduction}

Recently it has become fashionable to use derived categories
as a tool to study D-branes wrapped on complex submanifolds
of Calabi-Yau spaces.  Derived categories are now believed
to have a direct physical interpretation, via a number of
rather formal arguments.  (See \cite{paulron,medc,mikedc,paulalb} for
an incomplete list of early references on this subject.)

One prediction of this derived categories program is that
massless states of open strings between D-branes
wrapped on complex submanifolds of Calabi-Yau spaces should
be related to certain mathematical objects known as Ext groups.
For those readers not familiar with such technology,
Ext groups are analogous to cohomology groups, and are defined with
respect to two coherent sheaves.  The usual notation is
\begin{displaymath}
\mbox{Ext}^n_X\left( {\cal S}_1, {\cal S}_2 \right)
\end{displaymath}
where ${\cal S}_1$, ${\cal S}_2$ are two coherent sheaves on $X$
and $n$ is an integer.  Phrased in this language, if
we have one D-brane wrapped on a complex submanifold $i: S \hookrightarrow
X$ with holomorphic vector bundle ${\cal E}$ on $S$ and another D-brane
wrapped on a complex submanifold $j: T \hookrightarrow X$ with
holomorphic vector bundle ${\cal F}$ on $T$, then the prediction
in question is that massless states of open strings between these
D-branes should be counted by groups denoted
\begin{displaymath}
\mbox{Ext}^*_X\left( i_* {\cal E}, j_* {\cal F} \right)
\end{displaymath}
and
\begin{displaymath}
\mbox{Ext}^*_X\left( j_* {\cal F}, i_* {\cal E} \right)
\end{displaymath}
(depending upon the orientation of the open string).

This mathematically natural
prediction has been checked in a number of special cases.
For example, in the trivial case that both branes are wrapped on
the entire Calabi-Yau, the result is easily checked to be true.
See \cite{dfr} for a discussion of the
special case $\widetilde{{\bf C}^3/{\bf Z}_3}$.
Also see \cite{dgjt} for a self-consistency test of this
hypothesis in the special case of the quintic.
Additional special cases have also been checked \cite{dougprivcomm}.

Also note this prediction is closely analogous to some well-known
results in heterotic compactifications.  In heterotic compactifications
involving gauge sheaves that are not bundles \cite{dgm}, it has been
shown \cite{me,ralph} that massless modes are counted by Ext groups,
replacing the sheaf cohomology groups that count the massless modes
when the gauge sheaves are honest bundles \cite{distlergreene}.
Similarly, in the trivial case that the wrapped D-branes are wrapped
on the entire Calabi-Yau, the Ext groups reduce to sheaf cohomology.
The prediction that states are counted by Ext groups is equivalent
to the statement that for more general brane configurations than
the trivial one,
sheaf cohomology is replaced by Ext groups, which is certainly
what happened in heterotic compactifications.

However, there has not, to our knowledge, been any systematic
attempt to check directly in BCFT that open string states between
appropriate D-branes are always related to Ext groups.
In particular, the general correspondence between BCFT vertex operators
and Ext group elements does not exist in the literature.
Moreover, were it not for the fact that this classification of 
massless states is a prediction of a currently fashionable research
program, the claim might sound somewhat suspicious.  For example,
typically relations between physics and algebraic geometry rely
crucially on supersymmetry.  
Yet, the proposed
classification of massless states in terms of Ext groups
is needed to hold for non-BPS
brane configurations, as well as BPS configurations. 

In this paper, we shall begin to fill this gap in the literature.
Using standard well-known methods, we explicitly compute,
from first-principles, the
spectrum of (BRST-invariant) vertex operators corresponding
to massless Ramond sector states in open strings connecting 
D-branes wrapped on complex submanifolds of Calabi-Yau spaces
at large radius, and explicitly relate those vertex operators
to appropriate Ext group elements, for all possible configurations
of complex submanifolds (both BPS and non-BPS).

Although the methods involved are standard, we do find
some interesting physical subtleties in the open string case.
In the closed string case, such vertex operators are in
one-to-one correspondence with bundle-valued differential
forms, and so the spectrum of BRST-invariant vertex operators
is expressed in terms of a cohomology theory of such
bundle-valued differential forms, known as sheaf
cohomology\footnote{
Sheaf cohomology is defined for more general sheaves than
merely bundles.  Only in the special case that the
sheaves in question are locally-free, {\it i.e.} that they
correspond to bundles, does sheaf cohomology have a 
de-Rham-type description in terms of differential forms.
We will only be interested in sheaf cohomology valued in
bundles, not more general sheaves, and so for the purposes of
making this paper more accessible to a physics audience,
we will not distinguish between 
cohomology theories of bundle-valued differential forms and
more general sheaf cohomology.
}.  
For example, in heterotic strings with holomorphic
gauge bundle ${\cal E}$,
some of the massless modes are counted
by \cite{distlergreene}
\begin{displaymath}
H^n\left( \, X, \, \Lambda^m {\cal E} \, \right).
\end{displaymath}
For another example, in the closed string B model \cite{edtft},
there is a one-to-one correspondence between (BRST-invariant)
vertex operators and the sheaf cohomology groups
\begin{displaymath}
H^n\left( \, X, \, \Lambda^m TX \, \right).
\end{displaymath}
In the case at hand, a naive analysis of the massless Ramond
sector states in such open strings yields a counting in
terms of sheaf cohomology groups, and not Ext groups.
Although we show that the sheaf cohomology groups in question are 
always related
mathematically to Ext groups via spectral sequences,
these spectral sequences are often nontrivial --
although sheaf cohomology can be used to determine Ext groups,
a given Ext group element need not be in one-to-one correspondence
with any sheaf cohomology group element.
A more careful analysis reveals a physical subtlety that
has the effect of realizing the spectral sequences physically
in terms of BRST cohomology, so the spectrum of
massless Ramond sector states is, in fact, in one-to-one
correspondence with Ext group elements.

In the process of working out correspondences to
Ext groups, we also run across some interesting
interplays with other physics.
For example, the Freed-Witten anomaly, together with another
open string B model anomaly, plays a crucial role
in understanding how Ext groups arise.

Most of the paper is organized into a set of
case-by-case studies of intersecting branes of
increasing complexity.  We begin in section~\ref{bmodelrev}
by reviewing the open string B model, and in particular,
describe two anomalies that will play an important role
in deriving Ext groups.  One anomaly is the analogue for
open strings of the statement that the closed string B model
is only well-defined on Calabi-Yau's.  The other anomaly
is the Freed-Witten anomaly, which tells us that the gauge
bundle on a D-brane worldvolume is twisted to a non-honest
bundle, whenever the normal bundle to the worldvolume does not
admit a Spin structure.

In section~\ref{parcoin}, we
discuss the relationship between open string
boundary Ramond sector states and Ext groups in the simplest
case, namely that in which the complex submanifolds on which
the branes are wrapped are the same submanifold.
The boundary states for this particular case already exist
in the literature, although their relationship to Ext groups
does not seem to have been previously discussed.
The boundary states are naively counted by certain sheaf
cohomology groups, which are related to Ext groups via a spectral
sequence.  We discuss the spectral sequence in detail,
and include an example in which this spectral sequence is nontrivial,
in the sense that the unsigned sum of the number of boundary
vertex operators is {\it not} the same as the unsigned sum
of the dimensions of the Ext groups.
We describe the physical subtlety that alters the boundary state
analysis, and show how, in fact, the spectral sequence is realized
physically in BRST cohomology.  Thus, we see explicitly that
there is a one-to-one correspondence between massless Ramond
sector states (properly counted) and Ext group elements.

In section~\ref{pardiff} we consider the relationship
between boundary vertex operators and Ext groups in the next
simplest case, namely when one submanifold is itself a submanifold
of the other:  $T \subseteq S$.  We discuss the boundary vertex
operators and the spectral sequence relating the vertex operators
to Ext groups.  
We conjecture that the spectral sequence is realized physically as a 
modification of BRST cohomology, as happened in the case of
parallel coincident branes.
We also examine a naive problem with Serre duality
that crops up when the line bundle $\Lambda^{top} N_{T/S}$ is
nontrivial, a puzzle that is resolved in the next section.
We also note in this section that 
the degree of the Ext group as it arises in algebraic geometry
can differ from the charge of the
vertex operators (as used to determine the type of resulting
massless fields).

In section~\ref{gencase} we consider the general case
of two intersecting complex submanifolds $S$ and $T$, which need
not be parallel.  After disposing with the technical complications
introduced by having branes at angles, we find boundary vertex
operators and spectral sequences that generalize the results
of sections~\ref{parcoin} and \ref{pardiff}.  However,
in the general case there is an extremely interesting complication
that did not appear previously.  Previously there was always 
a spectral sequence relating the boundary vertex operators
to Ext groups.  However, in the general case, we find that
in order for such a relationship to exist, we must take into
account the Freed-Witten anomaly, which has been ignored in
previous treatments of sheaf models and derived categories.
This anomaly resolves apparent
difficulties with Serre duality (including the difficulty
first seen in section~\ref{pardiff}), 
as we discuss extensively.

In section~\ref{nonint} we very briefly dispose of the
case of nonintersecting branes.
In section~\ref{cpxes} we briefly begin to describe how
one can see Ext groups of complexes, not just individual torsion 
sheaves, in a special simple case.  (More extensive effort
will be delayed to later publications.)
Finally, in appendix~\ref{specseqap} we give mathematical derivations
of the spectral sequences that play an important role in the text.

In passing, note that we are primarily concerned with
only writing the spectrum of massless Ramond sector open string states
in a more elegant fashion.  Such analysis does not require
the target brane worldvolume theory to be well-behaved;
all we are doing is calculating part of the tree-level open
string spectrum.  For example, if the brane worldvolume theory
contains a tachyon, then it is unstable; however, as we are merely
rewriting the spectrum of string tree-level
massless Ramond sector boundary states, such tachyons would not
affect our calculations.  Similarly, our calculations are insensitive
to any anomalies in the target worldvolume theory.  Again, we are
merely rewriting part of the string tree-level open string spectrum;
whether the target worldvolume theory has tachyons or anomalies
certainly has a tremendous impact on the resulting physics,
but does not alter the string tree-level open string spectrum.

In the remainder of this paper, we shall
make the following assumptions.  First, all calculations
are performed at large-radius (closely analogous to the
original heterotic vertex operators calculated in
\cite{distlergreene}).  Second, we only consider branes
wrapped on (smooth) complex submanifolds of a Calabi-Yau,
whose intersections are again smooth submanifolds.
Thus, we are interested in counting massless Ramond sector
states, or equivalently, B-twisted topological field theory
states on the boundary of the open string.
Third, we shall assume throughout this paper that the
$B$ field vanishes identically.  Nonzero $B$ fields
play an interesting and important role in D-branes.  If we turn on a
$B$ field, the mathematical analysis can be handled using derived 
categories of twisted sheaves.  Since
the complications introduced are not relevant to the main point of
this paper, we content ourselves to set the $B$ field to zero.
Fourth, we shall only consider cases in which the
gauge `sheaf' on the brane worldvolume is an honest bundle;
we shall not attempt to study more general sheaves on 
the worldvolume of the brane.
Finally, there are no antibranes in this paper, only
branes.

\section{Review of the open string B model}  \label{bmodelrev}

\subsection{Actions and boundary conditions}

Following the conventions of \cite{edtft}, 
the bulk B model action can be written in the form
\begin{equation}
\label{tftaction}
\frac{1}{2} g_{i \overline{\jmath} } \partial \phi^{i}  
\overline{\partial} \phi^{ \overline{\jmath} } \: + \:
\frac{1}{2} g_{i \overline{\jmath} } \partial \phi^{ \overline{\jmath} }
\overline{\partial} \phi^{i} \: + \:
i g_{i \overline{\jmath} } \psi_{-}^{ \overline{\jmath} }
D_z \psi_{-}^{i} \: + \:
i g_{ i \overline{\jmath} } \psi_{+}^{\overline{\jmath}} 
D_{ \overline{z} } \psi_{+}^i \: + \:
R_{i \overline{\imath} j \overline{\jmath} }
\psi_+^i \psi_+^{\overline{\imath}}
\psi_-^j \psi_-^{\overline{\jmath}}
\end{equation}
where 
\begin{eqnarray*}
\psi_{\pm}^{\overline{\imath}} & \in & 
\Gamma\left( \phi^* T^{0,1} X \right), \\
\psi_+^i & \in & \Gamma \left( K \otimes
\phi^* T^{1,0} X \right), \\
\psi_-^i & \in & \Gamma\left( \overline{K} \otimes
\phi^* T^{1,0} X \right)
\end{eqnarray*}
and with BRST transformations
\begin{eqnarray*}
\delta \phi^i & = & 0, \\
\delta \phi^{ \overline{\imath} } & = & i \alpha \left(
\psi_+^{\overline{\imath}} \: + \: \psi_-^{ \overline{\imath} } \right), \\
\delta \psi_+^i & = & - \alpha \partial \phi^i, \\
\delta \psi_+^{\overline{\imath}} & = & -i \alpha
\psi_-^{ \overline{\jmath}} \Gamma^{\overline{\imath}}_{ \overline{\jmath}
\overline{m} } \psi_+^{ \overline{m} }, \\
\delta \psi_-^i & = & - \alpha \overline{\partial} \phi^i, \\
\delta \psi_-^{ \overline{\imath} } & = & - i \alpha
\psi_+^{ \overline{\jmath} } \Gamma^{ \overline{\imath} }_{
\overline{\jmath} \overline{m} } \psi_-^{ \overline{m} }.
\end{eqnarray*}
Following \cite{edtft}, we define
\begin{eqnarray*}
\eta^{ \overline{\imath} } & = & \psi_+^{ \overline{\imath} } \: + \:
\psi_-^{ \overline{\imath} }, \\
\theta_i & = & g_{ i \overline{ \jmath} } \, \left(
\psi_+^{ \overline{\jmath}} \: - \: \psi_-^{\overline{\jmath}}
\right), \\
\rho_z^i & = & \psi_+^i, \\
\rho_{ \overline{z} }^i & = & \psi_-^i,
\end{eqnarray*}
and it is easy to calculate that in the absence of background gauge fields,
the boundary conditions deduced from (\ref{tftaction}) are
\begin{displaymath}
\delta \eta^{ \overline{\imath}} \: = \:
\delta \theta_i \: = \: 0.
\end{displaymath}

Along Neumann directions,
\begin{displaymath}
\psi_+^i|_{ \partial \Sigma} \: = \: \psi_-^i |_{ \partial \Sigma}
\end{displaymath}
so we see that $\theta_i = 0$
for $i$ an index along
a Neumann direction, and similarly, along Dirichlet directions,
\begin{displaymath}
\psi_+^i |_{ \partial \Sigma} \: = \:
- \psi_-^i |_{ \partial \Sigma} 
\end{displaymath}
so $\eta^{\overline{\imath}} = 0$ for $i$ an index along Dirichlet directions.

In writing the boundary conditions above, we have neglected
two important subtleties, one mathematical, and the other physical:
\begin{enumerate}
\item First, although as $C^{\infty}$ bundles $TX|_S \cong TS \oplus N_{S/X}$
globally on $S$, as holomorphic bundles $TX|_S \not\cong TS \oplus N_{S/X}$
in general.  On any one (sufficiently small) complex-analytic local
neighborhood $U$ one can find complex-analytic coordinates 
such that $TX|_S|_U \cong TS|_U \oplus N_{S/X}|_U$, and such a choice
of coordinates is implicit in writing the local-coordinate expressions
for the boundary conditions given above.
However, because $TX|_S$ does not split globally on $S$, it is not
quite correct to say that $\theta_i = 0$ for directions ``normal'' to $S$
implies that the $\theta$'s couple to $N_{S/X}$, as one would naively
believe.
We shall speak more about this bit of mathematics in section~\ref{subt1}, 
when it
becomes physically relevant.
\item A second subtlety arises from physics, and is due to the
fact that along  Neumann directions, the Chan-Paton factors twist
the boundary conditions (see {\it e.g.} \cite{abooetal}),
so that in fact $\theta_i = \left( \mbox{Tr } F_{ i \overline{\jmath}} 
\right) \eta^{
\overline{\jmath}}$.
We will (eventually)  see
that taking into account that Chan-Paton-induced twist has the
effect of physically realizing spectral sequences discussed below
in terms of
BRST cohomology, so that the massless Ramond sector states
are in one-to-one correspondence with Ext group elements.
\end{enumerate}

\subsection{Two anomalies}   \label{twoanom}

Before proceeding to calculations of massless boundary Ramond spectra,
we should review two types of anomalies in the open string B model.

\subsubsection{Open string analogue of the Calabi-Yau condition}

The first anomaly we shall discuss is a close relative of
a certain closed string B model anomaly.
Recall the closed string B model is only well-defined for Calabi-Yau
target spaces \cite{edtft}, unlike the A model.
The reason for this is well-definedness of the integral over fermion
zero modes.  For example, when the worldsheet is a ${\bf P}^1$ and the target
is a three-fold,
there are three $\psi_+^{\overline{\imath}}$ and
three $\psi_-^{\overline{\imath}}$ zero modes, and to make sense of
the integration
\begin{displaymath}
\epsilon_{ \overline{\imath} \overline{\jmath} \overline{k} }
\int d \psi^{\overline{\imath}} d \psi^{\overline{\jmath}}
d \psi^{\overline{k}}
\end{displaymath}
implicitly assumes the existence of a trivialization 
$\epsilon_{\overline{\imath} \overline{\jmath} \overline{k} }$.
But, such a trivialization is a nowhere-zero (anti)holomorphic
top-form, which exists if and only if the target is a Calabi-Yau.
 
There is an analogous issue in the open string B model, though the
form of the anomaly varies depending upon the D-branes.
Assume that the worldsheet is an infinite strip, 
with one side on submanifold $S$ and the other on submanifold $T$,
the gauge bundle on the D-brane
on each side of the strip is trivial (simplifying the boundary conditions),
and $TX|_S$ splits holomorphically as $TS \oplus N_{S/X}$ for
each D-brane.   We will also assume that the intersection $S \cap T$
is a manifold.  Then, there are fermion zero modes coupling to
$T(S \cap T)$ and to 
\begin{displaymath}
\tilde{N} \: = \:
\frac{ TX|_{S \cap T} }{ TS|_{S\cap T} + TT|_{S \cap T} }
\end{displaymath}
(see the section on general intersections for more details).
Thus, because of these zero modes, the partition function is a section
of
\begin{displaymath}
\Lambda^{top} T^*(S \cap T) \otimes \Lambda^{top} \tilde{N}^{\vee}
\end{displaymath}
which (as we shall demonstrate in more detail later in 
section~\ref{restoreserre})
is isomorphic to
\begin{displaymath}
\Lambda^{top} N_{S \cap T/ S} \otimes \Lambda^{top} N_{S \cap T / T}
\end{displaymath}
Thus, in order for theory on the strip to be well-defined,
the line bundle
\begin{displaymath}
\Lambda^{top} N_{S \cap T / S } \otimes
\Lambda^{top} N_{S \cap T / T }
\end{displaymath}
must also be trivializable, 
so that the fermion zero mode integral is well-defined.

We conjecture that the case of more general boundary conditions can be
understood as arising from the determinant of the complex
\[
0\to T(S\cap T) \to T(S)|_{S\cap T}\oplus T(T)|_{S\cap T} \to T(X)|_{S\cap T}
\to 0
\]
which exists even if the normal bundle does not split.

We shall discuss this anomaly further in section~\ref{newselection},
where we shall check that it does not exclude any known
supersymmetric brane configurations, and also discuss how it
gives a new selection rule.

\subsubsection{The Freed-Witten anomaly}

The second class of anomalies that is relevant to this paper is due to 
Freed-Witten \cite{freeded}.
In their analysis\footnote{Although their paper was originally
written for physical untwisted open string theories,
the results also apply to the open string B model
\cite{freedpriv}.} of open string theories, they found two
interesting physical effects:
\begin{enumerate}
\item First, they found that a D-brane can only consistently wrap
submanifolds $S$ with the property that the normal bundle $N_{S/X}$
admits a Spin$^c$ structure.
\item Second, if the normal bundle $N_{S/X}$ admits a Spin$^c$ structure,
but not a Spin structure, then the gauge bundle on the D-brane worldvolume
must be twisted.
\end{enumerate}
All complex vector bundles admit Spin$^c$ structures, so the
first effect is irrelevant for our purposes.
The second effect is much more relevant, as not all complex vector
bundles\footnote{For example, the tangent bundle to the projective
plane ${\bf P}^2$ does not admit a Spin structure.
More generally, any complex vector bundle with $c_1$ odd does not admit
a Spin structure.} admit Spin structures.
It means that the gauge bundle on the D-brane worldvolume is not always
an honest bundle.  In particular, on the D-brane corresponding to the
sheaf $i_* {\cal E}$, the gauge bundle can not be merely ${\cal E}$.

This second effect might seem rather confusing, in light of the
fact that we usually identify sheaves with D-branes in a very direct way.
What this effect tells us is that the precise relationship between
sheaves and D-branes is slightly more subtle than we usually believe.
We need to work out the correct identification between physical branes
and sheaves.

Before we state the result, we want to emphasize an important point:
this identification cannot be unique.  The reason is that the derived
category of sheaves has autoequivalences.  In particular, if $L$ is
any line bundle on $X$, then tensoring with $L$ gives an
autoequivalence of the derived category.  So any identification
between branes and sheaves can only be well defined up to an overall
tensoring with $L$, or more precisely, its restiction to $S$.

Now, what is a correct way to take into account this twisting?
We will show that this can be done by
replacing the D-brane worldvolume bundle ${\cal E}$,
above, with the `bundle' 
${\cal E} \otimes \sqrt{ K_S^{\vee} }$, which is often not
an honest bundle, but rather a twisted bundle, in the sense
of \cite{freeded}.  This twisting is referred to as the
canonical Spin$^c$ lift in the literature.  
In particular, for $S$ a submanifold of a 
Calabi-Yau, $\sqrt{K_S^{\vee}}$ is an honest bundle if and only
if the normal bundle $N_{S/X}$ admits a Spin structure, not just
a Spin$^c$ structure, so we see that this ansatz does correctly
twist the worldvolume gauge bundle as prescribed in \cite{freeded}.
By the remarks above, it would have worked just as well to identify
sheaves with branes via associating to any bundle ${\cal E}$ on any $S$
the `bundle' ${\cal E} \otimes \sqrt{ K_S^{\vee}} \otimes L|_S$.

In Section~\ref{correct} we will show that this identification between
branes and sheaves identifies open string spectra with Ext groups of
sheaves in general. It appears to be the unique way to do so up to the
ambiguities discussed above.

Let us say a few words about this last point.  For fixed $S$, the most
general way to identify bundles on $S$ with branes is by fixing a
bundle $L_S$ and associating to the bundle ${\cal E}$ the `bundle'
${\cal E} \otimes \sqrt{ K_S^{\vee}} \otimes L_S$.  Our computations
in Section~\ref{correct} imply that if we look at open string spectra
with boundary conditions on different submanifolds $S$ and $T$, then
the spectra coincide with Ext groups if and only if 
\begin{equation}
\label{twistcond}
L_S|_{S\cap T}
=L_T|_{S\cap T}.
\end{equation}
Now let $L=L_X$.  Then (\ref{twistcond}) with $T=X$ says that $L_S =
L|_S$, and we are reduced to precisely the ambiguity noted above.  So
our identification of branes with sheaves is as unique as it can be.

Despite the ambiguity, it is both natural and convenient to fix it by
using the canonical Spin$^c$ lift, i.e.\ associating ${\cal E} \otimes
\sqrt{ K_S^{\vee} }$ to ${\cal E}$, and we will do so in the remainder of
this paper.

This choice is also the right one for describing D-brane charge in
terms of sheaves.  The ABS construction gives a commutative diagram
\begin{displaymath}
\xymatrix{
K(S) \ar[rr] \ar[dr]_{ \otimes \sqrt{ K_S^{\vee} } } & & K(X)\\
 & K^{tw}(S) \ar[ur] & 
}
\end{displaymath}
where the map from $K(S)$ to $K(X)$ is defined using the canonical
Spin$^c$ lift of $N_{S/X}$ and the canonical Spin$^c$ structure on $S$ is used
to identify twisted sheaves with twisted K-theory classes.  
One consequence of commutativity is that 
$\chi(E)$ equals the index of the Dirac operator of $E\otimes
\sqrt{K_S^\vee}$, which we will justify momentarily.  The index of
the Dirac operator is an invariant of the total D brane charge.  
Thus our identification
equates $\chi(E)$ with said invariant of the D brane charge.  
This makes quantitative the
conservation law observed in \cite{freeded}.

To justify the identification, we use the splitting principle and
formally write $c_1(S) = \sum_{i=1}^{\dim S} t_i$ and $c(E)=
\prod_{j=1}^{\mathrm{rank E}}(1+e_j)$.  Then by Riemann-Roch
\begin{displaymath}
\begin{array}{ccl}
\chi(E) &=& \int_S \mbox{ch}(E) \wedge \mbox{Td}(S)\\
        &=& \int_S \left(\sum_j \exp(e_j)\right)
            \prod_i \frac{t_i}{1-e^{-t_i}}
\end{array}
\end{displaymath}
while the invariant of the D brane charge for our conventionally 
associated brane 
${\cal E} \otimes \sqrt{ K_S^{\vee} }$ is by the Atiyah-Singer index theorem
\begin{displaymath}
\begin{array}{ccl}
N_0&=&\int_S \mbox{ch}\left({\cal E} \otimes \sqrt{ K_S^{\vee} }\right)
\hat{A}(S)\\
&=&\int_S\left(\sum_j \mathrm{exp}(e_j+(\sum_i t_i)/2)\right)
\prod_i\frac{t_i/2}{\mathrm{sinh}\ t_i/2}.
\end{array}
\end{displaymath}
The equality $N_0=\chi(E)$ is readily checked using the identity
\[
\frac{t}{1-e^{-t}}=e^{t/2}\frac{t/2}{\mathrm{sinh}\ t/2}.
\]

In any event, we can now see how to take into account the Freed-Witten
twisting.  {}From the discussion above, for a D-brane wrapped on a
submanifold $i: S \hookrightarrow X$, the worldvolume gauge bundle
that corresponds to the sheaf $i_* {\cal E}$ is given by ${\cal E}
\otimes \sqrt{K_S^{\vee}}$, and not ${\cal E}$.  This gauge bundle is
an honest bundle whenever the normal bundle admits a Spin lift, and is
not an honest bundle, otherwise.  We shall see later in
section~\ref{correct} how this particular twisting is uniquely
determined up to an overall line bundle by consistency with other
aspects of physics.

For many parts of this paper, we shall be able to simply
ignore this twisting by $\sqrt{K_S^{\vee}}$.
For example, when computing spectra between D-branes on the same
submanifold, each Chan-Paton factor will come with a $\sqrt{K_S^{\vee}}$,
and these factors will cancel one another out.
When considering D-branes wrapped on distinct submanifolds,
on the other hand, these factors will become extremely important,
and in fact we shall see that their presence is absolutely required
in order for Serre duality to close the spectra back into themselves,
and in fact to recover Ext groups at all.
Thus, for most of this paper we shall ignore the $\sqrt{K_S^{\vee}}$
twisting, and will only return to this issue in the section on
general intersections, where it will play a crucial role.

\section{Parallel coincident branes on $S \hookrightarrow X$}
\label{parcoin}

In this section we shall compute the massless Ramond sector spectrum
of 
open strings between two D-branes on the same complex
submanifold $S$ of a Calabi-Yau manifold $X$,
with inclusion $i: S \hookrightarrow X$.
We shall assume one of the branes has gauge fields
described by a holomorphic bundle ${\cal E}$,
and the other has gauge fields described by a 
holomorphic bundle ${\cal F}$.
Our methods are standard and well-known in the literature;
see for example 
\cite{distlergreene} for a closely related computation
of massless states in heterotic string compactifications
and
\cite{edtft} for another closely related computation
of vertex operators in the closed string B model.

\subsection{Basic analysis of massless boundary Ramond spectra}

Now, let us explicitly construct 
massless Ramond sector states,
assuming for the moment that $TX|_S$ splits holomorphically
as $TS \oplus N_{S/X}$, and that the Chan-Paton factors have no
curvature, so that the boundary conditions on the worldsheet fermions
are easy to discuss. 
These are states that, in an infinite strip, would
be placed in the infinite past, or alternatively,
if one conformally maps to an upper half plane with different
boundary conditions for $x > 0$ and $x < 0$, these are vertex
operators that would be placed on the boundary at $x=0$.
The calculational method we shall use is a simple extrapolation
of Born-Oppenheimer-based methods discussed in,
for example, \cite{distlergreene,edtft}.
Since we are working in a Ramond sector, the worldsheet bosons and
fermions contribute equally and oppositely to the normal ordering
constant, so massless states are constructed by acting on the
vacuum with zero modes.  Also, since we are dealing with
zero modes of strings, the Chan-Paton factors appear as nothing
more than indices on the vertex operators, as in \cite{edcs}.

For D-branes wrapped on the same complex submanifold $S \hookrightarrow X$,
as discussed above, we have boundary vertex operators
\begin{displaymath}
b^{\alpha \beta j_1 \cdots j_m}_{\overline{\imath}_1 \cdots 
\overline{\imath}_n}(\phi_0) \, \eta^{ \overline{\imath}_1} \cdots
\eta^{ \overline{\imath}_n} \theta_{j_1} \cdots \theta_{j_m}
\end{displaymath}
(where $\alpha$, $\beta$ are Chan-Paton indices).
Because of the boundary conditions,
the $\theta$ indices are constrained to only live along
directions normal to $S$, and the $\eta$ indices are constrained
to only live along directions tangent to $S$.
Also, because of boundary conditions the $\phi$ zero modes
$\phi_0$ are constrained to only map out $S$.
Also note that we are implicitly using $\theta$ and $\eta$ to denote
zero modes of both fields.
These vertex operators are in one-to-one correspondence with
bundle-valued differential forms living on $S$, and their BRST 
cohomology classes are identified with 
the (sheaf cohomology) group
\begin{equation}   \label{pcbranes}
H^n \left(S, {\cal E}^{\vee} \otimes {\cal F} \otimes 
\Lambda^m N_{S/X} \right)
\end{equation}
where $N_{S/X}$ is the normal bundle to $S$ in $X$.

Note in passing that this calculation is very similar
to several other closed string calculations, where analogous
results are obtained.  For example, closely related computations
in heterotic string compactifications with holomorphic
gauge bundle ${\cal E}$ show there are massless states counted by
\cite[section 3]{distlergreene}
\begin{displaymath}
H^n\left( \, X, \, \Lambda^m {\cal E} \, \right),
\end{displaymath}
and
in the closed string B model \cite{edtft}, vertex operators
are counted by the sheaf cohomology groups
\begin{displaymath}
H^n\left( \, X, \, \Lambda^m TX \, \right).
\end{displaymath}
Readers not familiar with the techniques being used may find
sheaf cohomology unfamiliar, but in fact sheaf cohomology is
nearly ubiquitous in these sorts of vertex operator computations.

The boundary states we have described above
are not new to this paper; the same vertex operators are 
also described in, for example, \cite[section 6.4]{brunneretal}
or more recently \cite{lmw}.
However, neither vertex operators for more general
brane configurations (in which both sides of the open
strings are not on the same submanifold) nor the relationship
of these vertex operators to Ext groups have been discussed
previously in the literature, and these topics will occupy the
bulk of our attention in this paper.

We should also take a moment to speak to potential boundary corrections
to the BRST operator.  In the vertex operator analysis above,
we implicitly assumed that the BRST operator on the boundary is the
same as the restriction of the bulk BRST operator to the boundary.
We claim that, modulo covariantizations, this is a reasonable assumption.
Two general remarks should be made to clarify this matter further.
\begin{itemize}
\item
First, from the Chan-Paton terms \cite{edcs}
\begin{displaymath}
\int \left( \phi^* A \: - \: i \eta^{\overline{\imath}} F_{
\overline{\imath} j } \rho^j \right)
\end{displaymath}
we find that the Noether charge associated with the BRST operator
picks up a term proportional to $A_{ \overline{\imath} } \eta^{
\overline{\imath}}$, where $A$ is the Chan-Paton gauge field.
This term merely serves to covariantize the BRST operator.
After all, the BRST operator essentially acts as $\overline{\partial}$,
but for fields coupling to bundles, one must add a connection term.
Thus, adding contributions from the Chan-Paton action to the
Noether current for the boundary BRST operator merely serves to
covariantize the BRST operator.
\item Second, in \cite[section 2.4]{paulalb}, 
certain additional boundary-specific
terms added to the BRST operator played an important role.
These terms arose after deforming the action, modelling giving a nonzero
vacuum expectation value to a tachyon in a brane-antibrane system.
Here, at no point will we consider deformations of the action.
Thus, no boundary-specific contributions to the BRST operator
of the form used in \cite{paulalb} will appear here.
\end{itemize}
Thus, in our analysis, the boundary BRST operator will always
be the restriction of the bulk operator to the boundary
(modified by covariantization with respect to the Chan-Paton gauge fields).

Serre duality acts to swap open string states of the
form~(\ref{pcbranes}) with those of open strings of
the opposite orientation.  To see this, a useful identity
is, for any complex bundle ${\cal G}$,
\begin{equation}
\label{det}
\Lambda^n {\cal G} \: \cong \: \left( \Lambda^{r-n} {\cal G}^{\vee} \right)
\otimes
\left( \Lambda^r {\cal G} \right)
\end{equation}
where $r = \mbox{rank } {\cal G}$, so as $\Lambda^{top} N_{S/X} \cong
K_S$, we see that Serre duality implies
\begin{displaymath}
H^n \left(S, {\cal E}^{\vee} \otimes {\cal F} \otimes 
\Lambda^m N_{S/X} \right)
\: \cong \:
H^{s-n} \left(S, {\cal F}^{\vee} \otimes {\cal E} \otimes
\Lambda^{r-m} N_{S/X} \right)^*
\end{displaymath}
where $s = \mbox{dim } S$ and $r = \mbox{dim }N_{S/X}$.
Also note that the boundary operator of maximal charge
that corresponds to the
holomorphic top form $\omega_{i_1 \cdots i_n} d z^{i_1} \wedge \cdots
\wedge dz^{i_n}$ of the Calabi-Yau always exists in this case
(assuming ${\cal E}={\cal F}$ and suppressing Chan-Paton indices)
and is given simply by
\begin{displaymath}
\overline{\omega}_{ \overline{\imath}_1 \cdots \overline{\imath}_s}^{
\: \: \: j_{s+1} \cdots j_n } \, \eta^{\overline{\imath}_1} \cdots
\eta^{\overline{\imath}_s} \theta_{j_{s+1}} \cdots \theta_{j_n}
\end{displaymath}
where $s = \mbox{dim } S$ and in this one equation
$n = \mbox{dim }X$.
Later, when considering more general boundary conditions,
we shall find cases in which Serre duality is no longer an involution
of the boundary vertex operator spectrum, and in such cases,
a maximal-charge vertex operator corresponding to the holomorphic
top form of the Calabi-Yau will no longer exist.

In the literature, it is frequently asserted that
open string modes are in one-to-one correspondence with
global Ext groups between torsion sheaves representing the
D-branes.  In the present case, this would be the claim
that the open string modes are in one-to-one correspondence
with elements of 
\begin{displaymath}
\mbox{Ext}^p_X \left( i_* {\cal E}, i_* {\cal F} \right)
\end{displaymath}
where $i_* {\cal E}$ and $i_* {\cal F}$ are sheaves supported
on $S \hookrightarrow X$, identically zero away from $S$,
that look like the bundles ${\cal E}$ and ${\cal F}$ over $S$.

By contrast to the assertions quoted above, our naive description
of the open string boundary vertex operators is in terms of 
bundle-valued differential forms which lead to (\ref{pcbranes})
rather than Ext groups.
However, that is not to say
they are unrelated to Ext groups; they do determine 
Ext groups mathematically via the spectral sequence
\begin{equation}
\label{easyss}
E_2^{p,q}:
H^p \left(S, {\cal E}^{\vee} \otimes {\cal F} \otimes \Lambda^q N_{S/X} 
\right) \: \Longrightarrow \:
\mbox{Ext}^{p+q}_X\left( i_* {\cal E}, i_* {\cal F} \right)
\end{equation}
(See appendix~\ref{specseqap} for a derivation.)

Earlier we mentioned that our boundary conditions were
slightly oversimplified, in that along Neumann directions,
the $\theta_i$ do not vanish, but rather obey
$\theta_i = \left(\mbox{Tr } F_{ i \overline{\jmath}} \right) \eta^{
\overline{\jmath}}$ \cite{abooetal}, something we have so far neglected.
In the special case that
the simpler boundary conditions are correct,
the spectral sequence above trivializes, and so
the sheaf cohomology groups are the same as Ext groups:
\begin{displaymath}
\mbox{Ext}^{n}_X\left( i_* {\cal E}, i_* {\cal F} \right) \: \cong \:
\bigoplus_{p+q=n} H^p\left(S, {\cal E}^{\vee} \otimes {\cal F}
\otimes \Lambda^q N_{S/X} \right)
\end{displaymath}
When the boundary conditions are more complicated,
we will find that the spectral sequence above is
realized physically via BRST cohomology.
In any event, we shall see that in all cases,
the massless Ramond sector states are actually in one-to-one 
correspondence with
Ext group elements.

For the moment, we shall check 
our vertex operator analysis in some simple examples in which
the spectral sequence is trivial.  After that,
we shall work through the subtleties in the boundary
conditions mentioned above.

\subsection{Examples}

We shall check our vertex operator counting in the following
two extreme cases: 
\begin{enumerate}
\item Branes wrapping a Calabi-Yau
\item Points on Calabi-Yau manifolds 
\end{enumerate}

First, consider branes wrapping an entire Calabi-Yau,
{\it i.e.}, $S = X$.  In this case, $N_{S/X} = 0$,
so the spectral sequence degenerates to give
\begin{displaymath}
Ext^n_X \left( {\cal E}, {\cal F} \right) \: = \:
H^n \left(X, {\cal E}^{\vee} \otimes {\cal F} \right)
\end{displaymath}
and the massless Ramond sector boundary states
are of the form
\begin{displaymath}
b^{\alpha \beta}_{ \overline{\imath}_1 \cdots \overline{\imath}_n }
\eta^{ \overline{\imath}_1 } \cdots \eta^{ \overline{\imath}_n }.
\end{displaymath}
These boundary states are well-known (see for example 
\cite{brunneretal}), and their low-energy interpretation is
a function of their $U(1)$ charge.  For example, from the
charge zero operator $b^{\alpha \beta}(\phi_0)$
one can construct a conformal dimension one operator
$\exp(- \phi) b^{\alpha \beta} \psi^{\mu}$,
where $\phi$ is the bosonized superconformal ghost and $\psi^{\mu}$
a worldsheet fermion transforming as a spacetime vector.
Such charge zero operators correspond in this fashion to
low-energy gauge fields.  A charge one operator, say
$b^{\alpha \beta}_{\overline{\imath}} \eta^{\overline{\imath}}$,
corresponds to a low-energy spacetime scalar,
with vertex operator of the form $\exp(- \phi) b^{\alpha \beta}_{
\overline{\imath}} \eta^{\overline{\imath}}$, which is a conformal
weight one operator transforming as a spacetime scalar.
As this story is well-known, we shall not belabor the point further.

Next, we shall consider branes `wrapped' on points
on Calabi-Yau threefolds.  Consider for example
$N$ D3-branes at a point $S$ on a Calabi-Yau threefold $X$.
For notational brevity, define ${\cal E} = {\cal O}^{\oplus N}$.
The nonzero sheaf cohomology groups are
\begin{displaymath}
\begin{array}{c}
H^0 \left(S, {\cal E}^{\vee} \otimes {\cal E} \right) \: = \:
{\bf C}^{N^2}, \\
H^0 \left(S, {\cal E}^{\vee} \otimes {\cal E} \otimes
N_{S/X} \right) \: = \: {\bf C}^{3 N^2}, \\
H^0 \left(S, {\cal E}^{\vee} \otimes {\cal E} \otimes
\Lambda^2 N_{S/X} \right) \: = \:
{\bf C}^{3 N^2}, \\
H^0 \left(S, {\cal E}^{\vee} \otimes {\cal E} \otimes
\Lambda^3 N_{S/X} \right) \: = \: {\bf C}^{N^2} 
\end{array}
\end{displaymath}
determining
\begin{displaymath}
\mbox{Ext}^n_X\left( i_* {\cal E}, i_* {\cal E} \right) \: = \:
\left\{ \begin{array}{ll}
{\bf C}^{N^2} & n = 0,3, \\
{\bf C}^{3 N^2} & n=1,2.
\end{array} \right.
\end{displaymath}
The first and last sheaf cohomology groups are Serre dual, and correspond to
open strings of opposite orientation; the second and third groups
are also Serre dual.
Thus, we need only consider the first two groups.
The first group describes states of $U(1)$ charge zero,
and so correspond in the low-energy theory to components of
a $U(N)$ gauge
field.
The second describes states of $U(1)$ charge one,
and so correspond in the low-energy theory to three
adjoint-valued fields.  Thus, we recover the expected field
content for D3-branes at a point on a Calabi-Yau threefold.

In these examples there was a natural correspondence
between the degree of the Ext group and the
$U(1)$ charge, as correlated with the type of
low-energy field (vector, scalar, {\it etc}).
However, in later sections we shall see explicitly
that unfortunately this correspondence cannot hold in general.

\subsection{First subtlety:  mathematics}  \label{subt1}

We mentioned earlier that there were two subtleties in the
boundary states.  The first subtlety described is,
on its face, an obscure mathematical point.
Namely, although for $C^{\infty}$ bundles, $TX|_S \cong
TS \oplus N_{S/X}$, this is not true for holomorphic bundles
in general.  As a result, the interpretation of boundary
conditions such as $\theta_i = 0$ is somewhat subtle.

We will see that this subtlety, on its own, has little
real effect.  Its proper understanding does not alter the
naive conclusion above, that massless Ramond sector states
appear to be counted by sheaf cohomology groups, and
not Ext groups.  In order to see explicitly that the
massless Ramond sectors states are actually counted by Ext groups,
we shall have to use the second subtlety mentioned.
However, although this subtlety will not have a significant
impact on the results,
its proper understanding will play a significant
role in the physical realization of the spectral sequence
discussed earlier, and for that reason we shall discuss it in detail.  

In general, globally on $S$, $TX|_S$ is merely an extension of
$N_{S/X}$ by $TS$:
\begin{displaymath}
0 \: \longrightarrow \: TS \: \longrightarrow \:
TX|_S \: \longrightarrow \: N_{S/X} \: \longrightarrow \: 0.
\end{displaymath}
One simple example in which $TX|_S$ does not split involves
conics $C$ in ${\bf P}^2$.  There,  
$TC = {\cal O}(2)$, $T {\bf P}^2|_C = {\cal O}(3) \oplus
{\cal O}(3)$, and $N_{C/{\bf P}^2} = {\cal O}(4)$, so clearly
$T {\bf P}^2 |_C \not\cong N_{C/{\bf P}^2} \oplus TC$; rather,
$T {\bf P}^2|_C$ is merely an extension of ${\cal O}(4)$ by ${\cal O}(2)$.
Although globally one cannot split $TX|_S$ holomorphically,
in any one sufficiently small complex-analytic local coordinate
patch, one can arrange for $TX|_S$ to split, and boundary
conditions such as $\theta_i = 0$ are implicitly written
in such special coordinates.

What effect does this subtlety have?
First, note that if we are working in a special case in which
$TX|_S$ {\it does} split, {\it i.e.}, in special cases in which
$TX|_S = N_{S/X} \oplus TS$ holomorphically globally on $S$,
then the analysis of the previous section goes through without a hitch.
If $TX|_S$ splits globally on $S$, then the local-coordinate expression
$\theta_i = 0$ does indeed imply that the $\theta$'s couple to
$N_{S/X}$, and the previous analysis is unchanged.

If $TX|_S \not\cong TS \oplus N_{S/X}$, then the analysis is
more complicated, but the result is the same.
For simplicity let us consider vertex operators with a single $\theta$,
naively corresponding to sheaf cohomology valued in $N_{S/X}$.
Since $TX|_S \not\cong TS \oplus N_{S/X}$, it is no longer true
that $\theta_i = 0$ implies that the $\theta$ couple to $N_{S/X}$.
After all, under a change of coordinates, $N_{S/X}$ will mix with
$TS$, and so the condition $\theta_i = 0$ for Neumann directions is
will not be invariant under holomorphic coordinate changes.
Rather, the $\theta$ merely couple to $TX|_S$, but are constrained
such that in certain special complex-analytic local coordinates,
some of the $\theta$ vanish.
In particular, sheaf cohomology valued in $N_{S/X}$ can no longer
be translated directly into vertex operators.

We can deal with this more complicated scenario as follows.
Although sheaf cohomology valued in $N_{S/X}$ can not be used
to write down vertex operators immediately, we can lift differential
forms valued in $N_{S/X}$ to differential forms valued in
$TX|_S$, and we {\it can} write down vertex operators associated
to those $TX|_S$-valued differential forms, since the $\theta_i$
couple to $TX|_S$.

Now, something interesting happens when we demand BRST invariance
of those newly-minted $TX|_S$-valued forms;
namely, they need no longer be $\overline{\partial}$-closed\footnote{
In the special
case that the $TX|_S$ splits globally on $S$, 
{\it i.e.} $TX|_S = N_{S/X} \oplus
TS$, then it is possible to generate a closed $TX|_S$-valued
differential form from any closed $N_{S/X}$-valued differential form.
When $TX|_S$ does not so split, this is not always possible.},
yet they still define BRST-invariant states.

Mathematically, we are taking advantage of a commuting diagram
which we shall write schematically as
\begin{displaymath}
\label{schematic}
\xymatrix{
{\cal A}^{0,n}(TS) \ar[r] \ar[d]^{ \overline{\partial} } &
{\cal A}^{0,n}(TX|_S) \ar[r] \ar[d]^{ \overline{\partial} } &
{\cal A}^{0,n}(N_{S/X}) \ar[d]^{ \overline{\partial} } \\
{\cal A}^{0,n+1}(TS) \ar[r] &
{\cal A}^{0,n+1}(TX|_S) \ar[r] &
{\cal A}^{0,n+1}(N_{S/X}) 
}
\end{displaymath}
where ${\cal A}^{0,n}$ denotes differential $(0,n)$ forms,
horizontal arrows are induced by the short
exact sequence above, the rows are exact, and we have  
suppressed the factors ${\cal E}^{\vee} \otimes {\cal F}$
throughout.
The image under $\overline{\partial}$ is a higher-degree $TX|_S$-valued
form, and from commutativity of the diagram above, that higher-degree
form is the image of a $TS$-valued form.

Technically what we are doing is realizing the coboundary map
in the long exact sequence of sheaf cohomology
\begin{displaymath}
H^n\left(S, {\cal E}^{\vee} \otimes {\cal F} \otimes N_{S/X} \right)
\: \longrightarrow \:
H^{n+1}\left(S, {\cal E}^{\vee} \otimes {\cal F} \otimes TS \right)
\end{displaymath}
induced by the short exact sequence
\begin{displaymath}
0 \: \longrightarrow \: TS \: \longrightarrow \: TX|_S \:
\longrightarrow \: N_{S/X} \: \longrightarrow \: 0.
\end{displaymath}
In algebraic topology, such a map is known as the Bockstein homomorphism.
We started with a $N_{S/X}$-valued form, and created a $TS$-valued
form of higher degree.
The physical vertex operators are defined by the $TX|_S$-valued differential
forms appearing in the first intermediate step.

Now, how can this be BRST invariant, as claimed?
We started with $N_{S/X}$-valued sheaf cohomology, lifted
the coefficients to $TX|_S$ to create differential forms that we
could associate to vertex operators, and then argued that
$\overline{\partial}$ of those differential forms gives 
$\overline{\partial}$-closed
$TS$-valued differential forms of one higher degree.
But in order to be BRST invariant, our vertex operators
(associated to $TX|_S$-valued forms) must be annihilated by 
$\overline{\partial}$.

The answer is in the boundary conditions $\theta_i = 0$ (for
Neumann directions).  These boundary conditions annihilate
$TS$-valued forms.  Thus, since the image of our vertex operators
under $\overline{\partial}$ is $TS$-valued,
our vertex operators are closed under the BRST transformation.

Now, the reader might well ask, why we went to the trouble
of working through these details.  What we have concluded, after
considerable effort, is that even though $TX|_S \not\cong
TS \oplus N_{S/X}$ holomorphically on $S$, the massless Ramond
sector states are nevertheless counted by the sheaf
cohomology groups
\begin{displaymath}
H^n\left(S, {\cal E}^{\vee} \otimes {\cal F} \otimes \Lambda^m N_{S/X} \right)
\end{displaymath}
(at least so long as we are dealing with the boundary
condition described as $\theta_i = 0$ for Neumann directions).
The vertex operators are slightly more complicated to express
than one would have naively thought, but at the end of the day,
we do not seem to have learned anything significantly new.

The reason we went to this trouble is that this complication
will play an important role when unraveling the next subtlety,
involving the altered boundary condition $\theta_i =
\left( \mbox{Tr } F_{i \overline{\jmath}} \right) \eta^{\overline{\jmath}}$.
The coboundary map discussed in detail above will form half of the
differential of the spectral sequence.  As here, vertex operators
will be associated to $TX|_S$-valued differential forms created
by lifting $N_{S/X}$-valued $\overline{\partial}$-closed differential
forms, and the BRST operator will act as $\overline{\partial}$ on
those $TX|_S$-valued forms.  Just as here, the result will be a 
$TS$-valued form, at which point we can apply the boundary
condition on the $\theta$'s.  Unlike the present case, the boundary
condition will not annihilate any $TS$-valued $\theta$'s, so demanding
BRST-invariance of our $TX|_S$-valued forms will give an additional
condition that will be equivalent to being in the kernel of the
differential of the spectral sequence.

\subsection{Second subtlety:  physics}

The second subtlety we mentioned previously was that
along Neumann directions, in suitable local complex-analytic coordinates,
it is not true that $\theta_i = 0$, but rather
$\theta_i = \left( \mbox{Tr } F_{i \overline{\jmath} } \right)
\eta^{\overline{\jmath}}$.
We shall deal with this subtlety in this section.
We shall find that this subtlety effectively alters the BRST cohomology
in such a way that the spectral sequence discussed earlier is
realized directly in BRST cohomology.  Thus, the spectrum of massless
Ramond sector states is counted directly by Ext groups,
instead of sheaf cohomology.

We begin this section by discussing the nontriviality of
the spectral sequence, followed by a detailed discussion
of the differentials of the spectral sequence.
Finally, we describe explicitly how those differentials are
realized physically.

\subsubsection{Nontriviality of the spectral sequence}

We have argued that, after making a slight simplification
of the boundary conditions, massless Ramond sector states in 
open strings are in one-to-one correspondence with sheaf
cohomology groups, related to Ext groups via a spectral
sequence.  We shall argue shortly that this spectral sequence
is realized physically after taking into account the 
correct boundary conditions, but before we work through those
details, we shall discuss the spectral sequence in greater detail.

In particular, in this subsection
we shall discuss the nontriviality of the spectral sequence,
because if the spectral sequence were always trivial, then there would
be little point in worrying about it.
In general, spectral sequences lose information -- the fact
that the obvious bigrading structure of the sheaf cohomology
groups reduces to a unigraded structure is one indication of this
loss of information.  However, in the explicit examples we have
computed above, it was the case that the spectral sequence
was trivial, in the sense that the dimension of an Ext group
was the same as the sum of the dimensions of the sheaf cohomology
groups feeding into it.

If it were always the case that the spectral sequence were trivial,
{\it i.e.}, if it were always the case that the number of independent sheaf
cohomology group elements 
was the same as the number of independent Ext group elements,
then our point that formally spectral sequences lose information
would seem rather moot, and discussions of physical realizations
of the spectral sequence would be rather pointless.

However, in general, the spectral sequence relating the sheaf
cohomology groups to Ext groups is not trivial -- the number of
independent boundary vertex operators is not the same as the number of
Ext group generators.  Thus, the map from boundary vertex operators to
Ext group elements is not invertible, in the strongest sense of the
term.  On the other hand, the signed sum of dimensions of
sheaf cohomology groups 
will always be the same as the signed sum of the dimensions of the Ext
groups.  This will be the case in general, as $d_r^{p,q}$ maps
$E_2^{p,q}$ to $E_2^{p+r,q-r+1}$ and so will always increase the
charge of an operator by 1.  Thus the spectral sequence will always
cancel out vertex operators in pairs, with differing sign in the
index.

One example in which this spectral sequence is nontrivial is as
follows.  Let $X$ be a K3-fibered Calabi-Yau threefold, and
let $S$ be a smooth K3 fiber.  Assume further that $S$ contains a 
$C\simeq{\bf P}^1$ which is rigid in $X$, having normal bundle
$N_{C/X}\simeq{\cal O}_C(-1)\oplus{\cal O}_C(-1)$.  This typically
implies that the bundle ${\cal O}_S(C)$ itself does not deform to first order
as $S$
moves in the fibration, but let's add that as another explicit assumption.

Let ${\cal E} = {\cal F} =
{\cal O}_S(C)$.  We claim that this gives an example with a non-trivial 
spectral sequence.

First note that the sheaf $i_*{\cal O}_S(C)$ does not deform, not even
to first order.  To see this, first observe that the support $S$ of
$i_*{\cal O}_S(C)$ can only deform in the given fibration; but then we have
assumed that $i_*{\cal O}_S(C)$ does not deform in the fibration 
so the sheaf
$i_*{\cal O}_S(C)$ does not deform in $X$ in any way whatsoever.

Next we note that ${\rm Ext}^1(i_*{\cal O}_S(C),i_*{\cal O}_S(C))$ is the 
space of first order deformations of the sheaf $i_*{\cal O}_S(C)$, which we
have just shown is 0.

But the spectral sequence (\ref{easyss}) has the nontrivial terms
\begin{displaymath}
E_2^{0,1}=
H^0(S,{\cal E}^\vee\otimes {\cal F}\otimes N_{S/X}) = H^0(S,{\cal O}) = {\bf C}
\end{displaymath}
and
\begin{displaymath}
E_2^{2,0}=
H^2(S,{\cal E}^\vee\otimes {\cal F}) = H^2(S,{\cal O}) = {\bf C}.
\end{displaymath}

It is immediate to additionally check that $E_2^{0,0}=E_2^{2,1}={\bf C}$
and all other terms in the spectral sequence are 0.

The spectral sequence (\ref{easyss}) has a differential $d_2^{0,1}:
E_2^{0,1}\to E_2^{2,0}$ which we will argue is nontrivial.  Since 
$E_2^{p,q}=0$ for $p\ge3$, all the differentials $d_r^{p,q}$ vanish for
$r\ge 3$.  So $d_2^{0,1}$ is the only differential in the spectral sequence
that could be nonzero.  If it were zero, then we would compute
${\rm Ext}^1(i_*{\cal O}_S(C),i_*{\cal O}_S(C))\ne0$, a contradiction.  
So $d_2^{0,1}$
is nonzero.  In particular we have an open string mode corresponding
to an element of $H^0(S,N_{S/X})$ which does {\em not\/} parametrize a nonzero
Ext element.

As an explicit example, consider $X=P(1,1,2,2,2)[8]$ considered in
\cite{cdfkm}.  The K3 fibration comes from the map $X\to{\bf P}^1$
sending $(x_1,\ldots,x_5)$ to $(x_1,x_2)$, and the general K3 fiber is
identified with a degree 4 K3 hypersurface in ${\bf P}^3$.  While a
general degree 4 surface contains no lines, it was argued that if $X$
has general moduli, then exactly 640 of these degree 4 K3 surfaces
contain a line, and that these are rigid in $X$.  Furthermore, still
choosing $X$ to have general moduli, we can assume that the general K3
fiber has Picard number at least 1.  Recall \cite[Pp.\ 590--94]{gh} 
that the moduli space $M_2$ of quartic 
K3 surfaces with Picard number at least 2 is a generically
smooth divisor in the moduli space $M_1$ of quartic K3's with Picard number 
1.  Given $X$, we get a map $\psi_X:{\bf P}^1\to M_1$ sending a point to the K3
fiber it parameterizes.  It is easy to check that if $X$ is general, then
$\psi_X$ meets $M_2$ transversally at smooth points.  This is enough to
guarantee that ${\cal O}_S(C)$ does not deform to first order as $S$ moves
in the K3 fibration.

\subsubsection{Details of the differentials}
\label{details}

We have just argued that the spectral sequence relating
sheaf cohomology to Ext groups is nontrivial in general,
so it is very important to check that that spectral sequence
really is realized physically.
Before we describe the physical realization of the spectral
sequence, we will first describe the differentials in more
detail.

Consider the special case of an open string connecting
a D-brane to itself.  In this case, we have the same
Chan-Paton gauge field on either side of the open string.
In this case, the level two differential 
\begin{displaymath}
d_2: \: H^0\left(S, {\cal E}^{\vee} \otimes {\cal E} \otimes
N_{S/X}  \right)
\: \mapsto \:
H^2\left( S, {\cal E}^{\vee} \otimes {\cal E} \right)
\end{displaymath}
is realized mathematically by the composition
\begin{displaymath}
H^0\left( S, {\cal E}^{\vee} \otimes {\cal E} \otimes N_{S/X} \right)
\: \longrightarrow \:
H^1\left( S, {\cal E}^{\vee} \otimes {\cal E} \otimes TS \right)
\: \longrightarrow \:
H^2\left( S, {\cal E}^{\vee} \otimes {\cal E} \right).
\end{displaymath}

The first map in the composition is the coboundary map
\begin{equation}
\label{coboundary}
H^0\left( S, {\cal E}^{\vee} \otimes {\cal E} \otimes N_{S/X} \right)
\: \longrightarrow \:
H^1\left( S, {\cal E}^{\vee} \otimes {\cal E} \otimes TS \right)
\end{equation}
in the long exact sequence of sheaf cohomology induced by the tensor product
of the short exact sequence
\begin{equation}
\label{nbs}
0 \: \longrightarrow \: TS \: \longrightarrow \: TX|_S \:
\longrightarrow \: N_{S/X} \: \longrightarrow \: 0
\end{equation}
with ${\cal E}^{\vee} \otimes {\cal E}$.
We discussed this coboundary map in detail in section~\ref{subt1}.
Recall from section~\ref{subt1} this coboundary map vanishes
if $TX|_S \cong TS \oplus N_{S/X}$ globally on $S$;
it is only nontrivial if $TX|_S$ does not split globally.
Put another way, if $TX|_S$ splits globally, then the spectral
sequence is trivial.

The second map in the composition is much easier to describe.
It involves contracting the $TS$ indices on the trace of the
curvature form of the connection on the bundle.
In other words, if $\theta_i$ schematically indicates a $TS$ direction,
then the second map in $d_2$ involves the replacement
\begin{equation}   \label{repl}
\theta_i \: \mapsto \:
\left( \mbox{Tr } F_{ i \overline{\jmath} } \right)
\, d \overline{z}^{ \overline{\jmath}}.
\end{equation}
The close relationship between the expression above and
the altered boundary conditions induced as in \cite{abooetal} is
no accident, and forms the heart of the physical realization
of the spectral sequence.

In principle, the higher differentials are constructed from
the same ingredients.  For example, let us consider
\begin{displaymath}
d_3: \: E_3^{0,2} \: \longrightarrow \: E_3^{3,0}.
\end{displaymath}
Note that $E_3^{0,2}$ consists of the part of $E_2^{0,2} = H^0(S, 
{\cal E}^{\vee} \otimes {\cal E} \otimes \Lambda^2 N_{S/X})$
that is annihilated by $d_2$.
Consider the short exact sequence
\begin{equation}
\label{timesn}
0\to N_{S/X}\otimes TX\to N_{S/X}\otimes TX|_S\to N_{S/X}\otimes N_{S/X}\to 0
\end{equation}
obtained from (\ref{nbs}) by tensoring with $N_{S/X}$.
In this case, $d_2$ acts by combining the coboundary map of (\ref{timesn})
with the replacement~(\ref{repl}).  In other words, to see
the action of $d_2$,
lift the $\Lambda^2 N_{S/X}$-valued zero form to a $N_{S/X} \otimes
TX|_S$-valued form, then apply $\overline{\partial}$ and
commutativity of (\ref{schematic}) to get a 
$N_{S/X}\otimes TS$-valued $(0,1)$-form.  Finally apply~(\ref{repl})
to get a $N_{S/X}$-valued two-form.  Here we have viewed $\Lambda^2 N_{S/X}$
as the subbundle of antisymmetric elements of $N_{S/X}\otimes N_{S/X}$.
The part of $H^0(S, 
{\cal E}^{\vee} \otimes {\cal E} \otimes \Lambda^2 N_{S/X})$
in the kernel of this map is $E_3^{0,2}$.
The differential $d_3$ acts on $E_3^{0,2}$ by lifting
the $\Lambda^2 N_{S/X}$-valued form to a $\Lambda^2 TX|_S$-valued form,
applying $\overline{\partial}$, and contracting both of the resulting
$TS$ indices with the curvature, using~(\ref{repl}).  The resulting 
indices correspond
to $TS$ rather than merely $TX|_S$ by the assumption that our 
$\Lambda^2N_{S/X}$-valued section is in the kernel of $d_3$.

\subsubsection{Physical realization of the spectral sequence}

So far in our analysis of massless Ramond sector states,
we have assumed that along Neumann directions, $\theta_i = 0$.
However, strictly speaking this is only the case when one
has trivial Chan-Paton gauge fields.
As noted many years ago in {\it e.g.} \cite{abooetal},
in the presence of nontrivial Chan-Paton gauge fields,
Neumann boundary conditions are twisted by the curvature
of the gauge field.
For example, for worldsheet scalars, ordinarily the
Neumann boundary conditions state that
\begin{displaymath}
\partial_{n} X \: = \: 0
\end{displaymath}
where $n$ denotes the direction normal to the boundary.
If the Chan-Paton factors have nontrivial curvature,
this condition is modified to become
\begin{displaymath}
\partial_n X^{\mu} \: = \: \left( \mbox{Tr } F^{\mu}_{\: \nu} \right)
\: \partial_t X^{\nu}.
\end{displaymath}
Although this twisting of boundary conditions seems to have
been largely ignored in most discussions of the open string
B model, it has played an important role elsewhere in
physics recently (see {\it e.g.} \cite{strometal}).

Let us carefully consider how this modifies our analysis
of massless Ramond sector states.
Boundary conditions for worldsheet fields coupling to
directions normal to the brane are unchanged, as otherwise
it would be impossible to make sense of the Chan-Paton action.
Boundary conditions for worldsheet fields coupling to directions
parallel to the brane, however, are changed as above.
One still has constant bosonic maps, as the modified
boundary conditions above only couple to derivatives of the
worldsheet bosons.  The boundary conditions on the worldsheet
fermions can now be written in suitable local coordinates as
\begin{equation}   \label{modbc}
\theta_i \: = \: \left( \mbox{Tr } F_{ i \overline{\jmath} } \right)
\: \eta^{\overline{\jmath}}.
\end{equation}
In general, if the Chan-Paton factors on either side of
the open string are different, then these boundary conditions
will change the fermion moding, and hence change the fermion
zero-mode structure.  Put another way, the effect of these
boundary conditions is very closely analogous to the
effect of having branes at angles, as discussed in
\cite{bdl}.  

We shall only consider the special case that the
Chan-Paton gauge fields on either side of the open string
are identical, {\it i.e.}, that the open string is connecting
a D-brane to itself.  This corresponds to the `dipole string'
case discussed in \cite{abooetal}.
In this special case, one has the same fermion zero modes
as assumed previously, so the analysis is very similar to that
discussed so far, except that the $\theta_i$ no longer couple
to $N_{S/X}$.  Instead, because the $\theta_i$ parallel to the
brane can be nonzero, the $\theta_i$ merely couple to $TX|_S$.

As before, boundary vertex operators should be of the general form
\begin{displaymath}
b^{\alpha \beta j_1 \cdots j_m}_{\overline{\imath}_1 \cdots 
\overline{\imath}_n}(\phi_0) \, \eta^{ \overline{\imath}_1} \cdots
\eta^{ \overline{\imath}_n} \theta_{j_1} \cdots \theta_{j_m}
\end{displaymath}
(where $\alpha$, $\beta$ are Chan-Paton indices).
Now, however, there are subtleties in the interpretation.
The $\theta_i$ couple to $TX|_S$, not $N_{S/X}$, and 
in principle $\theta_i$ parallel to the brane are related to the
$\eta^{\overline{\jmath}}$ by the boundary conditions.
In the special case that $TX|_S \cong N_{S/X} \oplus TS$ holomorphically,
we can simply ignore the $\theta_i$ parallel to the brane,
use only $\theta_i$ normal to the brane to construct the vertex operators above,
and immediately recover a classification in terms of
sheaf cohomology.  In this same case, the spectral sequence
relating sheaf cohomology to Ext groups is trivial.
When the spectral sequence is nontrivial,
$TX|_S \not\cong N_{S/X} \oplus TS$, but rather is merely an extension
of $N_{S/X}$ by $TS$.  In this case, although locally in coordinate
patches one can distinguish $\theta_i$ parallel to the brane
from $\theta_i$ normal to the brane, globally along $S$ one cannot
make such a distinction.

Let us assume that $TX|_S  \not\cong N_{S/X} \oplus TS$, and work
out how to describe the vertex operators.
For simplicity, for the moment we shall only consider vertex operators
of the form 
\begin{equation}    \label{vo1}
b^{\alpha \beta j}(\phi_0) \theta_j.
\end{equation}
We shall argue that computing BRST cohomology is equivalent to
evaluating the spectral sequence.

First, because the $\theta_i$ couple to $TX|_S$ and not $N_{S/X}$,
we cannot associate elements of $H^0({\cal E}^{\vee} \otimes {\cal E}
\otimes N_{S/X})$ with the vertex operator~(\ref{vo1}).
Also, because the $\theta_i$ ``parallel'' to the brane 
(not a well-defined notion when $TX|_S$ does not split holomorphically)
are related to the $\eta^{\overline{\jmath}}$ by the curvature of the
Chan-Paton factors, we cannot merely claim that the BRST cohomology
is simply $H^0({\cal E}^{\vee} \otimes {\cal E} \otimes TX|_S)$.
Instead, let us proceed more carefully.
We can manufacture an operator that is `close' to being BRST-closed
by starting with an element of $H^0({\cal E}^{\vee} \otimes {\cal E}
\otimes N_{S/X})$, and lifting the coefficients to $TX|_S$
to recover a (not-necessarily-closed) differential form valued
in $TX|_S$.  We can associate such a differential form with
a vertex operator of the form~(\ref{vo1}).  Unfortunately, the resulting
vertex operator need not be BRST-closed, as the corresponding
differential form is not $\overline{\partial}$-invariant, and because
of the boundary conditions on some of the $\theta_i$.

Let us now work out the action of the BRST operator on this vertex operator.
So, act on the vertex operator with the BRST operator,
or equivalently, act on the zero form with $\overline{\partial}$,
to generate a closed one-form.  We now have a closed one-form
valued in $TX|_S$.  Even better -- exactly as in the discussion of
the coboundary map in section~\ref{subt1}, our closed
$TX|_S$-valued one-form is mathematically the image of a
closed $TS$-valued one-form.

In other words, after applying $\overline{\partial}$,
we can now apply the boundary conditions
$\theta_i = \left( \mbox{Tr } F_{i \overline{\jmath}} \right)
\eta^{\overline{\jmath}}$
in a fashion that makes sense globally on $S$.
After applying this contraction, and comparing to the explicit
description of $d_2$ of the spectral sequence from the previous 
section, we see that the action
of the BRST operator is the same as the action of $d_2$;
demanding that the vertex operator be BRST-closed is equivalent
to demanding that it lie in the kernel of $d_2$.

Thus, in this fashion we see that the spectral sequence 
relating sheaf cohomology
to Ext groups is encoded physically in the BRST cohomology.

A very careful reader might note that we have glossed over one
important point.  In section~\ref{twoanom}, we discussed how the
Freed-Witten anomaly tells us that the sheaf $i_* {\cal E}$
corresponds to a D-brane with worldvolume gauge bundle ${\cal E}
\otimes \sqrt{ K_S^{\vee} }$.  The Chan-Paton curvature mentioned
above is the curvature of the twisted bundle ${\cal E} \otimes \sqrt{
K_S^{\vee} }$, yet the curvature appearing in the evaluation map
inside $d_2$ is the curvature of the bundle ${\cal E}$.  This is
consistent for the following reason.  The curvature of ${\cal E}
\otimes \sqrt{ K_S^{\vee} }$ can be expressed as the curvature of
${\cal E}$, plus an extra term determined by the first Chern class of
$K_S$.  However, that extra term drops out of the $d_2$ computation.

This can be seen readily from our computation of $d_2$ in
Section~\ref{details}.  We have interpreted the $d_2$ term as an
obstruction, but the bundle $TS$ is unobstructed for any deformation
of $S$; that is, $TS$ deforms for free with any deformation of $S$.
So $d_2$ must vanish if the gauge bundle is $TS$.  Furthermore, a
general $d_2$ is given by multiplying a coboundary map like
(\ref{coboundary}) with the curvature of the gauge field.  But the
curvature of the twisted gauge field is equal to the curvature of the
original gauge field plus half of the curvature of $TS$, and we have just 
observed that multiplying by the curvature of $TS$ gives zero.  Thus, that
extra term is irrelevant for $d_2$, and so it does not matter whether
we use the curvature of ${\cal E}$, or the curvature of ${\cal E}
\otimes \sqrt{K_S^{\vee}}$.

In this section we have described how the spectral sequence
can be realized physically, in the special case that the
Chan-Paton gauge fields on either side of the open string are the same.
In principle, of course, one would also like to check that 
the spectral sequence is realized physically more generally.
We hope to address this in future work, as this seems extremely plausible.

\section{Parallel branes on submanifolds of different dimension}
\label{pardiff}

\subsection{Basic analysis}

For another class of examples, consider a set of branes
wrapped on $i: S \hookrightarrow X$, with gauge fields defined
by holomorphic bundle ${\cal E}$, and another set of branes
wrapped on $j: T \hookrightarrow S \hookrightarrow X$, 
with gauge fields determined
by bundle ${\cal F}$.
Following the same analysis as above, and ignoring the
twisting of \cite{abooetal},
we find boundary Ramond sector states given by
\begin{displaymath}
b^{\alpha \beta j_1 \cdots j_m}_{ 
\overline{\imath}_1 \cdots \overline{\imath}_n}(\phi_0)
\eta^{ \overline{\imath}_1 } \cdots
\eta^{ \overline{\imath}_n }
\theta_{j_1} \cdots \theta_{j_m}
\end{displaymath}
where the $\eta$ indices are tangent to $T$,
and the $\theta$ indices are normal to $S \hookrightarrow X$.
Note that since the fields must respect the boundary conditions
on either side of the operator, there can be no fields with
indices in $N_{T/S}$, as any such $\eta$ fermions would be
killed by boundary conditions on one side,
and any such $\theta$ fermions would be killed by
boundary conditions on the other side.
Equivalently, if we think about fermions on an infinite strip,
fermions with mixed Dirichlet, Neumann boundary conditions 
are half-integrally moded, and so cannot contribute to massless
modes in the Ramond sector (where the zero-point energy is already zero).
The only possible factors can come from fermions with only Neumann
or only Dirichlet boundary conditions; hence, the vertex operators
above couple to the tangent bundle of $T$ and to $N_{S/X}$, but not
$N_{T/S}$.
These vertex operators are in one-to-one correspondence with
bundle-valued differential forms, counted by the
(sheaf cohomology) groups
\begin{equation}  \label{pddspec}
H^n \left(T, ( {\cal E}|_T )^{\vee} \otimes {\cal F} \otimes
\Lambda^m N_{S/X} |_T
\right).
\end{equation}
Again, we are here ignoring the boundary condition twisting described 
in \cite{abooetal}.

We should mention that our expressions for sheaf cohomology
groups describing modes of open strings connecting parallel
branes of different dimensions do not assume that the difference
in dimensions is a multiple of four.
If the difference in dimensions is not a multiple of four,
then in a physical theory, supersymmetry is\footnote{Technically speaking
we are discussing `undissolved' branes, not `dissolved'
branes.  If the second brane is not really a second boundary
condition on the open string, but only merely curvature in the
Chan-Paton bundle, then of course the difference in `dimensions'
need not be a multiple of four.} broken -- although
the Ramond sector ground state will always have vanishing zero-point energy,
it is well-known that
only when the difference in dimensions is a multiple of four
will it be possible to find corresponding massless modes in the 
Neveu-Schwarz sector.
However, there are massless modes in the Ramond sector for any difference
in dimensions, and also corresponding BRST-invariant TFT states
for any difference in dimensions.

As in the last section, one would hope that these open string states
should be related to global Ext groups of the form
\begin{displaymath}
\mbox{Ext}^p_X\left( i_* {\cal E}, j_* {\cal F} \right)
\end{displaymath}
where $i: S \hookrightarrow X$ and $j: T \hookrightarrow X$ are
inclusion maps.  As before, we have a minor puzzle, in that the
open string states are not counted by such Ext groups,
but rather by the sheaf cohomology groups~(\ref{pddspec}).
As before, the resolution of this puzzle is that there is a spectral
sequence relating the sheaf cohomology groups~(\ref{pddspec}) counting
the open string vertex operators to the desired Ext groups.
Specifically, there is a spectral sequence generalizing
(\ref{easyss}) as follows:
\begin{equation}
\label{intss}
E_2^{p,q} \: = \: H^p \left(T, {\cal E}^{\vee}|_T \otimes {\cal F}
\otimes \Lambda^q N_{S/X}|_T \right) 
\: \Longrightarrow \:
\mbox{Ext}^{p+q}_X\left( i_* {\cal E}, j_* {\cal F} \right).
\end{equation}
(See appendix~\ref{specseqap} for a derivation.)

It is plausible to assume that, as in the last section,
when the Chan-Paton-induced boundary condition twisting
described in \cite{abooetal} is properly taken into account,
the effect will be to realize the spectral sequence above
physically in the BRST cohomology, so that the states of the
massless Ramond sector spectrum will be in one-to-one correspondence
with Ext group elements.  
We would like to
check this explicitly in future work.
For the rest of this section, we shall describe the vertex operators
in terms of sheaf cohomology groups, and leave explicit checks of the
physical realization of the spectral sequence above to future work.

Let us now look for an example where the spectral sequence
(\ref{intss}) is nontrivial and $T\ne S$.  If we are to have a nontrivial $d_r$
with $r\ge 2$ then clearly some 
\begin{displaymath}
E_2^{p,q}\: =\: H^p \left(T, {\cal
E}^{\vee}|_T \otimes {\cal F} \otimes \Lambda^q N_{S/X}|_T \right)
\end{displaymath}
with $p\ge 2$ must be nonzero.  In particular, it must be the case
that ${\rm dim}(T)\ge 2$.  Since $T$ is a proper subset of $S$, it
must be the case that ${\rm dim}(S)\ge 3$.  But if ${\rm dim}(S)={\rm
dim X}=3$, then $N_{S/X}=0$, hence $E_2^{p,q}=0$ for all $q\ge1$, and again
the spectral sequence must degenerate.  The conclusion is that 
nontrivial spectral sequences~(\ref{intss}) 
can occur only if ${\rm dim}(X)\ge 4$.

\subsection{Example:  ADHM construction}

As a quick check of our spectrum computation,
let us check that our description in terms
of sheaf cohomology groups correctly reproduces details
of the ADHM construction.  For $k$ $D5$-branes on $N$ $D9$-branes,
say, one should expect to recover a single six-dimensional
hypermultiplet valued in $(k,N)$ of $U(k) \times U(N)$,
and a single hypermultiplet valued in the adjoint of $U(k)$.
Assume $T$ is a point on a $K3$ surface $S = X$ (so $N_{S/X}$ is trivial),
and ${\cal E}$, ${\cal F}$ are both trivial, then the only nonzero sheaf
cohomology is
\begin{equation}
\label{dfn}
H^0\left({\rm pt}, ({\cal E}|_T)^{\vee} \otimes {\cal F} \right)
\: = \: {\bf C}^{kN}
\end{equation}
from open strings of one orientation between the $D5$ and $D9$ branes, 
determining
\begin{displaymath}
\mbox{Ext}^n_{K3}\left( i_* {\cal E}, j_* {\cal F} \right) \: = \:
\left\{ \begin{array}{ll}
{\bf C}^{kN} & n= 0, \\
0 & n \neq 0,
\end{array} \right.
\end{displaymath}
a well-known mathematical result, and also
\begin{equation}
\label{dnn}
H^0\left({\rm pt}, {\cal F}^{\vee} \otimes {\cal F} \right), \:
H^0\left({\rm pt}, {\cal F}^{\vee} \otimes {\cal F} \otimes N_{T/X} \right), \:
H^0\left({\rm pt}, {\cal F}^{\vee} \otimes {\cal F} \otimes 
\Lambda^2 N_{T/X} \right)
\end{equation}
from our previous analysis applied to strings connecting 
$D5$ branes to $D5$ branes.

Now, in a physical theory
these sheaf cohomology groups are counting massless
fermions, so from~(\ref{dfn}), we see 
we get a single
$(k,N)$-valued fermion in six dimensions.
This is precisely the fermionic content\footnote{
Recall that a single six-dimensional Weyl fermion is equivalent to a pair of
``symplectic-Majorana'' Weyl fermions, which allow us to
write the supersymmetry transformations in a form some readers
might find more familiar.  Put another way, in order to get
a pair of four-dimensional Weyl fermions after compactification
on $T^2$, one must have started with a single six-dimensional
Weyl fermion.
} 
of a six-dimensional hypermultiplet
valued in $(k,N)$ of $U(k) \times U(N)$.  The set of states~(\ref{dnn})
from open strings connecting $D5$ branes to $D5$ branes,
precisely describe the fermion content of the
six-dimensional gauge multiplet, a 
$U(k)$-adjoint-valued hypermultiplet, and their antiparticles.
(The antiparticles of the $D5-D9$ string states are given by
strings with opposite orientation, as we will be able to check later
after doing the general case which includes in particular
the situation $S\subset T$  arising when the opposite orientation is
chosen.)

Now, in closed string theories, the GSO projection uniquely
determines the type of low-energy field ({\it e.g.}, chiral multiplet
or vector multiplet) from the $U(1)$ charge of the vertex operator.
In particular, vertex operators of $U(1)$ charge one correspond
to low-energy chiral multiplets in compactifications to four
dimensions.  
With that in mind, it would be natural to assume that
the degree of the Ext group determines the type of low-energy field,
in the same way.  In other words, it would be natural to assume
that a field associated to a vertex operator corresponding
to an Ext group element of degree one (or whose Serre dual has
degree one) should correspond to a chiral multiplet, and so forth.

Unfortunately that naive assumption is not true in general,
as we see in this example.
Specifically, in this ADHM example the hypermultiplets are coming
from Ext groups of degree zero, not one.
(We are describing Ext groups on K3's, but the problem persists
even after performing the obvious further compactification
on $T^2$, as we shall check shortly.)
Thus,
we see explicitly in this ADHM example that such a hypothetical
correspondence between degrees of Ext groups and type of matter
content simply does not hold in general.
Here, our hypermultiplets are coming from Ext groups of
degree zero, not one.

Note that we have not used any results from this paper
in making this observation.
Also note we are not claiming that scalar fields are never associated
with Ext groups of degree one in nontrivial cases.  For example,
in the next section, we shall see
a nontrivial example in which the scalar fields are associated with Ext groups
of degree one.
Note furthermore that this ADHM example is not the only example
in which this naive mismatch occurs.  
For example, in work to be published
shortly we shall see that the same problem arises when describing
configurations of D5 branes and D9 branes on orbifolds, as relevant
to, for example,the ADHM/ALE construction.

There are several possible resolutions of this discrepancy.
One possibility is that the $U(1)$ charge of a state and
the degree of the Ext group do not match, perhaps via
(fractionally) charged vacua.
(This has also been suggested by others; see {\it e.g.}
\cite{mikedc}.)  Perhaps the states have $U(1)$ charge one,
and non-matching Ext degree.  Of course, it would be
absurd to then claim that the corresponding Ext groups are actually
of degree one just because of their $U(1)$ charges, as those
degrees are uniquely determined mathematically and, in fact, are
well-known.  
In this paper we have specifically avoided talking about
$U(1)$ charges of states, so as to avoid having to sort
out such issues.  
We do not intend to try to give a definitive account of the
resolution of this puzzle in this paper.

Instead, we merely wish to observe, based on this
very clean example, that in general the type of matter
content is obviously not determined solely by the degree
of the Ext group; just because an Ext group element is not
of degree one does not mean it cannot describe scalar states.
The correct statement is obviously more complicated.

Also note that if we compactify the $D5$ branes on a $T^2$,
then (\ref{dfn}) is replaced by the sheaf cohomology groups
\begin{eqnarray*}
H^0\left({T^2}, ({\cal E}|_T)^{\vee} \otimes {\cal F} \right) 
& = & {\bf C}^{kN}, \\
H^1\left({T^2}, ({\cal E}|_T)^{\vee} \otimes {\cal F} \right)
& = & {\bf C}^{kN}
\end{eqnarray*}
in one open string orientation,
corresponding to 
\begin{displaymath}
\mbox{Ext}^n_{K3 \times T^2}\left( i_* {\cal E}, j_* {\cal F} \right)
\: = \: 
\left\{ \begin{array}{ll}
{\bf C}^{kN} & n=0,1, \\
0 & n > 1,
\end{array} \right.
\end{displaymath}
respectively,
which give us two fermions in the $(k,N)$ of $U(k) \times U(N)$,
again precisely correct to match the fermionic content of
four-dimensional hypermultiplets.
Note in this case, when $X$ has complex dimension three
instead of two, one of the matter fields does come from
an Ext group of degree one, though neither the other, nor its
Serre dual, come from an Ext group of degree one.

Again, we shall not attempt in this paper 
to give a definitive account of the
relationship between degrees of Ext groups and type of matter field;
rather, we merely wish to point out that the relationship is
obviously rather more complicated than seems to be often assumed.

\subsection{Serre duality invariance of the spectrum}

In this section we shall point out a puzzle
involving Serre duality.  Ordinarily spectra are
Serre duality invariant, but in the present case,
we shall see that Serre duality invariance is naively
lost in certain cases.  We shall explore this naive loss of Serre duality
invariance in this section, and in a later section we shall
point out how Serre duality invariance is restored by
an interesting physical effect.

How does Serre duality act on our boundary vertex operators?
In general, we can use the relation\footnote{This can
be derived by taking determinants in the exact sequence $0\to T(T)\to
T(S)|_T\to N_{T/S}\to 0$, i.e.\ $(K_S)^\vee|_T\simeq
K_T^\vee\otimes \Lambda^{\rm top}N_{S/T}$.  It is also theorem III.7.11 in
\cite{hartshorne}, and when $T$ is complex codimension one
in $S$, this reduces to the adjunction formula
\cite[p. 147]{gh}.  We shall use analogous formulas repeatedly
in the rest of this paper, but will not give such detailed
justification in future.} 
$K_T \cong K_S|_T \otimes
\Lambda^{top} N_{T/S}$.
As a result, under Serre duality,
\begin{eqnarray*}
H^n \left(T, ( {\cal E}|_T )^{\vee} \otimes {\cal F} \otimes
\Lambda^m N_{S/X} |_T
\right)
 & \cong &
H^{t-n} \left(T, {\cal F}^{\vee} \otimes {\cal E}|_T \otimes
\Lambda^m N^{\vee}_{S/X} |_T \otimes K_T \right)^* \\
& \cong & H^{t-n} \left(T, {\cal F}^{\vee} \otimes {\cal E}|_T \otimes
\Lambda^{r-m} N_{S/X} |_T \otimes K^{\vee}_S|_T \otimes K_T \right)^* \\
 & \cong &
 H^{t-n} \left(T, {\cal F}^{\vee} \otimes {\cal E}|_T \otimes
\Lambda^{r-m} N_{S/X} |_T \otimes \Lambda^{top}N_{T/S} \right)^*
\end{eqnarray*}
where $t$ is the dimension of $T$ and $r$ is the codimension of $S$ in $X$.
In the second isomorphism we have used $\Lambda^rN_{S/X}\simeq K_S$.

This result is rather interesting, and somewhat
unexpected.  Ordinarily, the spectrum of string states is
invariant under Serre duality -- not only for the open string
boundary states for parallel coincident branes that we
discussed in the last section, but also in other contexts,
such as large-radius heterotic
compactifications \cite{distlergreene}.
By contrast, we seem to see here that the 
open string spectrum connecting parallel branes
of different dimensions  
is not invariant under Serre duality in general.

To shed a little more light on this subject, let us try to find
a maximal-charge boundary vertex operator corresponding to the
holomorphic top form on the Calabi-Yau.  Such operators
are deeply intertwined with Serre duality invariance of
spectra, and they play
important roles in ${\cal N}=2$ supersymmetry algebras.
For example, these operators are typically identified with spectral
flow by one unit; recall spectral flow by half a unit is
part of the spacetime supercharge.

As one might have expected by now,
we find that such maximal-charge boundary vertex operators
do not always exist.
We can write\footnote{As mentioned earlier, in general, the restriction of
the tangent bundle of X to a submanifold is merely an
extension of the normal bundle by the tangent bundle
and need not split,
so in general it need not be true that $TX|_T \cong T^*T \oplus N_{T/S}
\oplus N_{S/X}|_T$.
}
\begin{displaymath}
\Lambda^{top} T^* X \: \cong \:
\Lambda^{top} T^* T \otimes \Lambda^{top} N^{\vee}_{T/S} \otimes
\Lambda^{top} N^{\vee}_{S/X}|_T.
\end{displaymath}
If $\Lambda^{top} N_{T/S}$ is trivial,
then the holomorphic top form on the Calabi-Yau determines
a section $h$ of $\Lambda^{top} T^* T \otimes \Lambda^{top} N^{\vee}_{S/X}|_T$,
and so if ${\cal E}={\cal O}_S$, ${\cal F}={\cal O}_T$
we have a maximal-charge boundary vertex operator given by
\begin{displaymath}
\overline{h}_{\overline{\imath}_1 \cdots \overline{\imath}_t}^{
\: \: \: j_{t+1} \cdots j_n } \,
\eta^{ \overline{\imath}_1 } \cdots \eta^{ \overline{\imath}_t }
\theta_{j_{t+1}} \cdots \theta_{j_n}
\end{displaymath}
where $t = \mbox{dim } T$ and, in this one example,
$n = \mbox{dim }X$.
On the other hand, if $\Lambda^{top} N_{T/S}$ is not trivial,
then it is not clear that the holomorphic top-form on the Calabi-Yau
determines any boundary vertex operator.

Thus, whenever the line bundle $\Lambda^{top} N_{T/S}$ is
nontrivial, the spectrum of boundary vertex operators
appears to lose Serre duality
invariance, and there is a corresponding lack of
a maximal-charge boundary vertex operator induced by the
holomorphic top form of the Calabi-Yau.  
In the next section, when we discuss general
brane intersections, we shall return to this issue.
Specifically, we shall find that the presence of this line
bundle is a reflection of the Freed-Witten anomaly \cite{freeded}
discussed in section~\ref{twoanom}.
More to the point, were it not for the Freed-Witten anomaly,
spectra could not be Serre-duality invariant, and we would not
even be able to claim that spectra are counted by Ext groups.
We shall discuss this issue in more detail in the next section.

\section{General intersecting branes}  \label{gencase}

\subsection{Basic analysis}

Next, consider two branes, both wrapped on complex
submanifolds of a Calabi-Yau, intersecting nontrivially.
As before, we shall work out the spectrum of boundary 
Ramond sector states.
Also as before, since the Ramond sector vacuum has vanishing zero-point
energy, such ground states are guaranteed to exist,
regardless of whether the corresponding brane configuration
is supersymmetric.

In the general intersecting brane case we have the additional
complication that we must now treat branes intersecting at general angles.
Previously we have only considered parallel branes,
so all worldsheet fermions were either integrally
or half-integrally moded, depending upon boundary conditions.
For branes at general angles, fermions can be
fractionally\footnote{The actual calculation in
\cite{bdl} is more nearly appropriate to branes wrapped
on special Lagrangian submanifolds; however, it is easy to
repeat the analysis for branes on complex manifolds at angles,
and one recovers the same result that the moding is shifted.} 
moded \cite{bdl}, which naively would appear
to greatly complicate our calculations.

A moment's thought reveals that no great complication is introduced.
We are calculating Ramond sector ground states, and the
Ramond sector vacuum has vanishing zero-point energy, so there can
be no contribution from any fermions whose moding is non-integral.
Fractionally moded fermions are therefore irrelevant.

Following the same analysis as before, if $S$ and $T$ denote
two intersecting complex submanifolds of the Calabi-Yau $X$,
with inclusions $i$, $j$ respectively, and holomorphic
bundles ${\cal E}$, ${\cal F}$, respectively, 
such that their intersection $S \cap T$ is another
submanifold, then as before, if we assume that restrictions of tangent
bundles split holomorphically and that Chan-Paton factors have no curvature,
then
we find boundary states given by
\begin{equation}  \label{genstates}
b^{\alpha \beta j_1 \cdots j_m}_{ 
\overline{\imath}_1 \cdots \overline{\imath}_n}(\phi_0)
\eta^{ \overline{\imath}_1 } \cdots
\eta^{ \overline{\imath}_n }
\theta_{j_1} \cdots \theta_{j_m}
\end{equation}
where the $\phi$ zero modes describe sheaf cohomology
on the intersection $S \cap T$, the $\eta$ indices are
tangent to the intersection $S \cap T$, and the $\theta$ indices
are normal to both $S$ and $T$.
More formally, the $\theta$'s are sections of the bundle
\begin{equation}
\label{bundleb}
\tilde{N} \: = \: TX|_{S\cap T} / \left( \, TS|_{S \cap T} + TT|_{S \cap T} \, \right)
\end{equation}
defined on $S \cap T$,
so the  
boundary states above are in one-to-one correspondence
with elements of the sheaf cohomology groups
\begin{equation}   \label{try1}
H^n \left({S \cap T}, {\cal E}^{\vee}|_{S \cap T} \otimes
{\cal F}|_{S \cap T} \otimes \Lambda^m \tilde{N} \right).
\end{equation}
Proceeding as before, it would be natural
to conjecture the existence of a spectral sequence
\begin{equation}    \label{gencase1}
E_2^{p,q} \: = \:
H^p \left({S \cap T}, {\cal E}^{\vee}|_{S \cap T} \otimes
{\cal F}|_{S \cap T} \otimes \Lambda^q \tilde{N} \right)
\: \Longrightarrow \:
\mbox{Ext}^{p+q}_X \left( i_* {\cal E}, j_* {\cal F} \right).
\end{equation}
Unfortunately, we have a problem -- no such spectral sequence
exists in general, as we shall argue
in the next section.  
After we have demonstrated where our analysis has been slightly
too naive, we shall describe the physics subtlety that 
we have glossed over, and describe how to correctly count
both physical states, as well as the relation between
the correctly-counted physical states and Ext groups.

As before, we are also ignoring the Chan-Paton-induced
twisting of boundary conditions described in \cite{abooetal}.
Judging from the case of parallel coincident branes, it is
extremely plausible that all spectral sequences are realized
physically in BRST cohomology, but we are not at present
able to explicitly perform that check.
For the remainder of this section, we shall ignore the
Chan-Paton-induced boundary condition twisting, and leave
the study of its effects to future work.

\subsection{Failure of the naive analysis}   \label{fails}

The proposed spectral
sequence~(\ref{gencase1}) that would be needed to relate 
the proposed sheaf
cohomology groups in~(\ref{try1}) to the desired Ext groups
does not exist in general, as we shall now demonstrate.
Our counterexample consists of two complex submanifolds
$S$ and $T$, intersecting transversely in a point,
such that $S$ is a divisor in the ambient Calabi-Yau $X$.
Since these are transverse submanifolds intersecting in a
point, from the analysis above the only possible boundary
vertex operators are charge zero operators of the form
$b^{\alpha \beta}(\phi_0)$, corresponding to elements of
$H^0\left(S\cap T, {\cal E}^{\vee}|_{S \cap T} \otimes
{\cal F}|_{S \cap T} \right)$, and hence if the desired spectral
sequence existed in general, the only nonzero Ext group
would be $\mbox{Ext}^0_X\left( i_* {\cal E}, j_* {\cal F} \right)$.

Since $S$ is a divisor in $X$, we can calculate the Ext groups
directly.  For simplicity, assume that ${\cal E} = {\cal O}_S$
and ${\cal F} = {\cal O}_T$.
Without loss of generality assume $S$ is the zero locus of a section
of a line bundle ${\cal O}_X(S)$, then we have a projective resolution
of ${\cal O}_S$ given by
\begin{displaymath}
0 \: \longrightarrow \: {\cal O}_X(-S) \: \longrightarrow \:
{\cal O}_X \: \longrightarrow \: {\cal O}_S \: \longrightarrow \: 0.
\end{displaymath}
Local $\underline{\mbox{Ext}}^*_X\left( {\cal O}_S, {\cal O}_T \right)$
sheaves are given by the cohomology sheaves of the complex
\begin{displaymath}
\underline{\mbox{Hom}}_{ {\cal O}_X } \left( {\cal O}_X, {\cal O}_T \right)
\: \longrightarrow \: 
\underline{\mbox{Hom}}_{ {\cal O}_X } \left( {\cal O}_X(-S), {\cal O}_T
\right)
\end{displaymath}
which we can rewrite as
\begin{displaymath}
{\cal O}_T \: \longrightarrow \:
{\cal O}_T(S|_T).
\end{displaymath}
The map above is injective, with cokernel ${\cal O}_{S \cap T}\left(
S|_{S \cap T} \right) \cong N_{S/X}|_{S \cap T}$.
Thus, 
\begin{displaymath}
\underline{\mbox{Ext}}^n_{ {\cal O}_X } \left( {\cal O}_S, {\cal O}_T
\right) \: = \:
\left\{ \begin{array}{cl}
        N_{S/X}|_{S \cap T} & n = 1 \\
        0 & n \neq 1  \end{array} \right.
\end{displaymath}
so from the local-global spectral sequence we immediately compute that
\begin{displaymath}
\mbox{Ext}^n_X \left( {\cal O}_S, {\cal O}_T \right) \: = \:
\left\{ \begin{array}{cl}
        H^0 \left({ S \cap T }, N_{S/X}|_{S \cap T} \right) & n = 1 \\
        0 & n \neq 1 \end{array} \right.
\end{displaymath}
(using the fact that $S \cap T$ is a point)
but this contradicts the claim above, as here we see in this example
that the nonzero Ext groups all have degree one or greater, whereas in order
for our conjectured spectral sequence (\ref{gencase1}) to hold,
the only nonzero Ext group must be at degree zero.

\subsection{Corrected analysis}
\label{correct}

We have a puzzle.  Previously, in discussions of parallel branes,
we were able to relate boundary vertex operators to Ext groups
in a reasonably straightforward fashion that worked for
all parallel brane configurations, both BPS and non-BPS.
In the general case, however, after repeating the same analysis
as before, we find it is not possible to relate our
boundary vertex operators to Ext groups in the general case.
Given our previous success, we must surely have made an
error in our analysis.
But where?

The resolution of this puzzle lies in the fact that we have
neglected the Freed-Witten anomaly \cite{freeded}.
Recall from section~\ref{twoanom} that as a result of that anomaly,
the sheaf $i_* {\cal E}$ corresponds to a D-brane with bundle
${\cal E} \otimes \sqrt{ K_S^{\vee} }$ on its worldvolume,
instead of ${\cal E}$.
In the present case, that means that the states~(\ref{genstates})
are counted by sheaf cohomology groups on $S \cap T$ valued
in the bundle
\begin{displaymath}
\left. \left( {\cal E} \otimes \sqrt{ K_S^{\vee} } 
\right)^{\vee}\right|_{S \cap T}
\otimes
\left. \left( {\cal F} \otimes \sqrt{ K_T^{\vee} } \right)\right|_{S \cap T}
\otimes
\Lambda^m \tilde{N}
\end{displaymath}
In other words, because of the Freed-Witten anomaly there is a factor
of 
\begin{displaymath}
\sqrt{ \frac{ K_S|_{S \cap T} }{ K_T|_{S \cap T} } }
\end{displaymath}
that we missed previously.

Now, on the face of it, we do not seem to have improved matters
significantly.  After all, that square-root-bundle is not always an honest
bundle, and sheaf cohomology with coefficients in non-honest bundles
is not well-defined.

The other anomaly discussed in section~\ref{twoanom} saves the day.
Recall that just as the closed string B model is only well-defined for
Calabi-Yau targets, the open string B model is only well-defined when
the following\footnote{Technically this is the condition that applies when $TX|_S$ and $TX|_T$ split holomorphically and the Chan-Paton factors have
no curvature, so that the open string boundary conditions are easy.
This is the same set of conditions for relevant spectral sequences to
trivialize, so that Ext groups are the same as sheaf cohomology groups
we shall obtain shortly.  If these conditions are not met, then the open
string zero modes are more complicated, spectral sequences are nontrivial,
and the line bundle~(\ref{trivbdle}) 
is also modified.} line bundle is trivial:
\begin{equation}  \label{trivbdle}
\Lambda^{top} N_{S \cap T / S } \otimes \Lambda^{top} N_{S \cap T / T}
\end{equation}

Using the fact that
\begin{eqnarray*}
\Lambda^{top} N_{S \cap T / X } & \cong & \Lambda^{top} N_{S \cap T / S}
\otimes \Lambda^{top} N_{S/X}|_{S \cap T} \\
& \cong & \Lambda^{top} N_{S \cap T/ T} \otimes
\Lambda^{top} N_{T/X}|_{S \cap T}
\end{eqnarray*}
and the fact that $K_S = \Lambda^{top} N_{S/X}$,
we see that whenever the line bundle~(\ref{trivbdle}) is trivial,
{\it i.e.} whenever the open string B model is well-defined,
\begin{eqnarray*}
\sqrt{ \frac{K_S|_{S\cap T}}{K_T|_{S\cap T}} } & \cong &
\Lambda^{top} N_{S \cap T/T} \\
\sqrt{ \frac{K_T|_{S \cap T}}{K_S|_{S \cap T}} } & \cong &
\Lambda^{top} N_{S \cap T / S}
\end{eqnarray*}
so those square roots are actually honest bundles whenever the open
string B model is well-defined, and in fact the Freed-Witten anomaly yields
new factors in the coefficients of the sheaf cohomology groups.

In other words, taking into account the Freed-Witten anomaly,
we see that the boundary Ramond sector states~(\ref{genstates})
are 
in one-to-one correspondence
with elements of the sheaf cohomology groups
\begin{equation}  \label{try2}
\begin{array}{c}
H^p\left(S\cap T, {\cal E}^{\vee}|_{S \cap T} \otimes 
{\cal F}|_{S \cap T} \otimes \Lambda^{q-m} \tilde{N} \otimes
\Lambda^{top} N_{S \cap T/T } \right) \\
H^p\left(S\cap T, {\cal E}|_{S \cap T} \otimes 
{\cal F}^{\vee} |_{S \cap T} \otimes \Lambda^{q-n} \tilde{N}
\otimes \Lambda^{top} N_{S \cap T/S} \right)
\end{array}
\end{equation}
(depending upon open string orientation)
where $m = \mbox{rk } N_{S \cap T/T}$, $n = \mbox{rk } N_{S \cap T/S}$.

Unlike the attempt described above in~(\ref{try1}) to associate
sheaf cohomology groups with physical states,
our new sheaf cohomology groups above in~(\ref{try2})  
{\it are} related to Ext groups, via the spectral sequences below:
\begin{eqnarray*}
E_2^{p,q} \: = \: H^p\left(S\cap T, {\cal E}^{\vee}|_{S \cap T}
\otimes {\cal F}|_{S \cap T} \otimes \Lambda^{q-m} \tilde{N}
\otimes \Lambda^{top} N_{S \cap T / T} \right)
& \Longrightarrow &
\mbox{Ext}^{p+q}_X \left( i_* {\cal E}, j_* {\cal F} \right) \\
E_2^{p,q} \: = \: H^p\left(S\cap T, {\cal E}|_{S \cap T}
\otimes {\cal F}^{\vee}|_{S \cap T} \otimes
\Lambda^{q-n} \tilde{N} \otimes \Lambda^{top} N_{S \cap T / S } \right)
& \Longrightarrow &
\mbox{Ext}^{p+q}_X \left( j_* {\cal F} , i_* {\cal E} \right)
\end{eqnarray*}
where $m$ is the rank of $N_{S \cap T/T}$ and $n$ is the rank
of $N_{S \cap T/S}$.
Mathematical proofs of these spectral sequences
can be found in appendix~\ref{specseqap}.

Note that our example in subsection~\ref{fails} is fixed
by taking into account the Freed-Witten anomaly.  
Recall there we considered two branes
with trivial bundles wrapped on two transverse submanifolds $S$
and $T$, intersecting in a point, such that $S$ is a divisor
in the ambient Calabi-Yau.  From our new analysis~(\ref{try2}),
the possible boundary vertex operators are classified by
the single sheaf cohomology group
\begin{displaymath}
H^0\left(S\cap T, N_{S \cap T/T} \right)
\end{displaymath}
(for one open string orientation).
Using the fact that $\tilde{N} = N_{S/X}|_{S \cap T} / (N_{S \cap T/T})$
and that $\tilde{N} = 0$ to see that $N_{S \cap T/T} = N_{S/X}|_{S \cap T}$,
we find that the only physical states (in one orientation)
are naively counted by the sheaf cohomology group
\begin{displaymath}
H^0\left(S\cap T, N_{S/X}|_{S \cap T} \right)
\end{displaymath}
and so the corresponding Ext groups are
\begin{displaymath}
\mbox{Ext}^n_X \left( {\cal O}_S, {\cal O}_T \right)
\: = \: \left\{ \begin{array}{cl}
H^0\left( S \cap T, N_{S/X}|_{S \cap T} \right) & n = 1 \\
0 & n \neq 1
\end{array} \right.
\end{displaymath}
completely agreeing with the computations described in
section~\ref{fails}.

Let us check our results in another example.  
Consider (following \cite{dmflds}) a pair of sets of orthogonal D-branes
on ${\bf C}^3$, which we shall describe with
complex coordinates $x$, $y$, $z$.  Put
$N$ branes on the divisor $y = z = 0$ in ${\bf C}^3$
and $k$ branes on the divisor $x = y = 0$ in ${\bf C}^3$.
In \cite{dmflds}, it was claimed that open strings stretching
between these D-branes should form a hypermultiplet valued in the
$(k,N)$ of $U(k) \times U(N)$.  So, in order to agree,
the sheaf cohomology groups
for one orientation must be two copies of ${\bf C}^{kN}$
(as four-dimensional hypermultiplets
contain a pair of Weyl fermions).
If we take $S$ to be the worldvolume of the
first set of branes, and $T$ the worldvolume of the second set,
with ${\cal E}$ a trivial rank $N$ bundle on $S$ and ${\cal F}$ a
trivial rank $k$ bundle on $T$,
then we find that $TX|_{S \cap T} / \left(
TS|_{S \cap T} + TT|_{S \cap T} \right)$ is the trivial rank 1 complex
vector bundle
over $S \cap T$ ({\it i.e.}, the origin of ${\bf C}^3$), corresponding
to the directions $y$, $\overline{y}$, along which the
open string has Dirichlet boundary conditions on both sides.
Also, $N_{S\cap T/S}$ and $N_{S \cap T/T}$ are both
rank one trivial bundles over the point $S \cap T$ (the origin
of ${\bf C}^3$), and so we get two sheaf cohomology
groups in each orientation, namely 
\begin{displaymath}
\begin{array}{c}
H^0\left(S\cap T, {\cal E}^{\vee}|_{S \cap T} \otimes
{\cal F}|_{S \cap T} \otimes N_{S \cap T/T} \right) 
\: = \: {\bf C}^{kN}, \\
H^0\left(S\cap T, {\cal E}^{\vee}|_{S \cap T} \otimes
{\cal F}|_{S \cap T} \otimes \tilde{N} \otimes N_{S \cap T/T}
\right) \: = \: {\bf C}^{kN}
\end{array}
\end{displaymath}
for one orientation, determining
\begin{displaymath}
\mbox{Ext}^n_{{\bf C}^3}\left( i_* {\cal E}, j_* {\cal F} \right)
\: = \: 
\left\{ \begin{array}{ll}
{\bf C}^{kN} & n=1,2, \\
0 & n \neq 1,2.
\end{array} \right.
\end{displaymath} 
This is the correct number of states to give a four-dimensional
hypermultiplet valued in the $(k,N)$ of $U(k) \times U(N)$,
precisely reproducing the result in \cite{dmflds}.
Note that in this case, each Ext group (or its Serre dual) corresponding
to a matter field has degree one, agreeing with current lore,
unlike the ADHM example discussed
previously.

As another check, we shall rederive from our (corrected)
general analysis our results
for the case of parallel branes of different dimension.
Suppose that $T$ is a submanifold of $S$.
Then, in the expressions above, $\tilde{N} = TX|_{S \cap T} / \left(
TS|_{S \cap T} + TT|_{S \cap T} \right) = N_{S/X}|_T$, 
$N_{S \cap T / T} = 0$, and $N_{S \cap T/S} = N_{T/S}$ in this
case.
Thus, the two spectral sequences (\ref{try2}) reduce to
\begin{eqnarray*}
E_2^{p,q} \: = \: H^p \left(T, {\cal E}^{\vee}|_T \otimes {\cal F}
\otimes \Lambda^q N_{S/X}|_T \right) 
& \Longrightarrow &
\mbox{Ext}^{p+q}_X \left( i_* {\cal E}, j_* {\cal F} \right) \\
E_2^{p,q} \: = \: H^p \left(T, {\cal E}|_T \otimes {\cal F}^{\vee}
\otimes \Lambda^{q-n} N_{S/X}|_T \otimes \Lambda^n N_{T/S} \right)
& \Longrightarrow & 
\mbox{Ext}^{p+q}_X \left( j_* {\cal F}, i_* {\cal E} \right)
\end{eqnarray*}
for $n = \mbox{rk } N_{T/S}$.
The first of these expressions is the first spectral sequence
we discussed in describing how to generate Ext groups from
boundary vertex operators for parallel branes of different dimension,
and the second we discussed later in that section in connection
with Serre duality in non-supersymmetric cases.

Note in passing that there are two families of cases in which
the spectral sequences below~(\ref{try2}) completely degenerate, and
Ext groups can be identified canonically with single
sheaf cohomology groups:
\begin{enumerate}
\item Suppose $S$ and $T$ intersect transversely.
In this case, $\tilde{N} = 0$ as\footnote{For transversely
intersecting submanifolds, this is usually stated
for the tangent bundles as $C^{\infty}$ real vector bundles;
however, it is also true for the associated holomorphic
vector bundles we have here.} $TS + TT = TX$ over $S\cap T$,
so $E_2^{p,q} = 0$ if $q \neq \mbox{rk } N_{S \cap T/T}$ in the first
spectral sequence, and $E_2^{p,q} = 0$ if $q \neq \mbox{rk } N_{S \cap T/S}$
in the second.  Hence, the spectral sequences
completely degenerate, and
\begin{eqnarray*}
\mbox{Ext}^p_X \left( i_* {\cal E}, j_* {\cal F} \right) & = & 
\left\{ \begin{array}{cl}
H^{p-m} \left({S \cap T}, {\cal E}^{\vee}|_{S \cap T} \otimes
{\cal F}|_{S \cap T} \otimes \Lambda^{top} N_{S \cap T/T} \right) & p \geq m 
\\
0 & p < m   \end{array}  \right. \\
\mbox{Ext}^p_X \left( j_* {\cal F}, i_* {\cal E} \right) & = &
\left\{ \begin{array}{cl}
H^{p-n} \left({S \cap T}, {\cal E}|_{S \cap T} \otimes
{\cal F}^{\vee}|_{S \cap T} \otimes \Lambda^{top} N_{S \cap T/S}
\right) & p \geq n 
\\
0 & p < n  \end{array} \right.
\end{eqnarray*}
where $m = \mbox{rk } N_{S \cap T/T}$ and
$n =  \mbox{rk } N_{S \cap T/S}$.
\item Suppose $S \cap T$ is zero-dimensional.
In this case, $E_2^{p,q} = 0$ for $p \neq 0$ in both
spectral sequences, and so we find
\begin{eqnarray*}
\mbox{Ext}^p_X \left( i_* {\cal E}, j_* {\cal F} \right) & = &
\left\{ \begin{array}{cl}
H^0 \left({S \cap T}, {\cal E}^{\vee}|_{S \cap T} \otimes
{\cal F}|_{S \cap T} \otimes
\Lambda^{p-m} \tilde{N} \otimes \Lambda^{top} N_{S \cap T/T} \right)
& p \geq m \\ 
0 & p < m  \end{array} \right. \\
\mbox{Ext}^p_X \left( j_* {\cal F}, i_* {\cal E} \right) & = &
\left\{ \begin{array}{cl}
H^0 \left({S \cap T}, {\cal E}|_{S \cap T} \otimes {\cal F}^{\vee}|_{
S \cap T} \otimes \Lambda^{p-n} \tilde{N} \otimes \Lambda^{top} N_{S \cap T/S}
\right) & p \geq n \\ 
0 & p < n \end{array} \right.
\end{eqnarray*}
with $m$ and $n$ as above.
\end{enumerate}

\subsection{Restoration of Serre duality invariance}  \label{restoreserre}

In section~\ref{pardiff} we saw a breakdown in
Serre duality invariance of the open string spectra.
However, at the time we did not take into account
the possibility that the boundary vacua could be
sections of bundles over part of the Calabi-Yau.
In this section, we shall see explicitly that by taking into
account the Freed-Witten anomaly,
Serre duality invariance of the spectrum is restored.

How does Serre duality act on our states?
The sheaf cohomology groups
\begin{equation}   \label{genserre}
H^p \left({S \cap T}, {\cal E}^{\vee}|_{S \cap T} \otimes
{\cal F}|_{S \cap T} \otimes \Lambda^{q-m} \tilde{N} \otimes \Lambda^m
N_{S \cap T/T} \right) 
\end{equation}
are isomorphic to
\begin{displaymath}
H^{s-p} \left({S \cap T}, {\cal E}|_{S \cap T} \otimes
{\cal F}^{\vee}|_{S \cap T} \otimes \Lambda^{b-q+m} \tilde{N}
\otimes \Lambda^{top} \tilde{N}^{\vee} \otimes \Lambda^m N_{S \cap T/T}^{\vee}
\otimes K_{S \cap T} \right)^*
\end{displaymath}
where $b = \mbox{rk } \tilde{N}$ and $s = \mbox{dim }S \cap T$.
Next, use the fact (to be demonstrated below) that
\begin{displaymath}
\Lambda^{top} \tilde{N}^{\vee} \otimes \Lambda^{top} N^{\vee}_{S \cap T/T}
\: \cong \:
\Lambda^{top} N_{S \cap T/S} \otimes \Lambda^{top}N_{S \cap T/X}^{\vee}
\end{displaymath}
so that the sheaf cohomology groups~(\ref{genserre})  are isomorphic to
\begin{displaymath}
H^{s-p} \left({S\cap T}, {\cal E}|_{S \cap T} \otimes
{\cal F}^\vee|_{S \cap T} \otimes \Lambda^{b-q+m} \tilde{N}
\otimes \Lambda^{top} N_{S \cap T/S}
\otimes \Lambda^{top} N_{S \cap T/X}^{\vee} \otimes K_{S \cap T}
\right)^*
\end{displaymath}
but $K_{S \cap T} \cong \Lambda^{top} N_{S \cap T/X}$, so we finally
see that the sheaf cohomology groups~(\ref{genserre}) are isomorphic to
\begin{displaymath}
H^{s-p} \left({S\cap T}, {\cal E}|_{S \cap T} \otimes
{\cal F}^\vee|_{S \cap T} \otimes \Lambda^{b-q+m} \tilde{N}
\otimes \Lambda^{top} N_{S \cap T/S}
\right)^*.
\end{displaymath}
Thus, Serre duality acts to exchange the sheaf cohomology groups
appearing in our two spectral sequences.
In other words, taking into account the Freed-Witten anomaly,
we find that the physical
spectrum is Serre duality invariant.

Let us also determine
under what circumstances the holomorphic top form on the
ambient Calabi-Yau induces a maximal-charge boundary vertex operator.
Proceeding as before, 
we have that
\begin{displaymath}
\Lambda^{top} T^* X|_{S \cap T} \: = \:
\Lambda^{top} T^* (S \cap T) \otimes \Lambda^{top} N_{S \cap T / X}^{\vee}.
\end{displaymath}
Next, use the fact that
\begin{eqnarray*}
\Lambda^{top} N_{S \cap T/X} & = &
\Lambda^{top} N_{S \cap T/T} \otimes \Lambda^{top} N_{T/X}|_{S \cap T} \\
& = & \Lambda^{top} N_{S \cap T/S} \otimes \Lambda^{top}
N_{S/X}|_{S \cap T}
\end{eqnarray*}
and as
\begin{displaymath}
\tilde{N} \: = \: \frac{TX |_{S \cap T} }{ TS|_{S \cap T} + TT|_{S \cap T} }
\: = \:
\frac{ N_{S/X}|_{S \cap T} }{ N_{S \cap T/T} } \: = \:
\frac{ N_{T/X}|_{S \cap T} }{ N_{S \cap T/S} }
\end{displaymath}
we see that
\begin{displaymath}
\Lambda^{top} N_{S \cap T/X} \: = \:
\Lambda^{top} N_{S \cap T/T} \otimes 
\Lambda^{top} \tilde{N} \otimes
\Lambda^{top} N_{S \cap T/S }
\end{displaymath}
so finally
\begin{displaymath}
\Lambda^{top} T^* X|_{S \cap T} \: = \:
\Lambda^{top} T^* (S \cap T) \otimes
\Lambda^{top} N_{S\cap T/T}^{\vee} \otimes
\Lambda^{top} N_{S \cap T/S}^{\vee} \otimes
\Lambda^{top} \tilde{N}^{\vee}.
\end{displaymath}
Thus, if both $\Lambda^{top} N_{S \cap T/S}$ and 
$\Lambda^{top} N_{S \cap T/T}$ are trivial,
then the holomorphic top form on the ambient Calabi-Yau is
equivalent to a section of $\Lambda^{top} T^* (S \cap T)
\otimes \Lambda^{top} \tilde{N}^\vee$, which is equivalent to a 
maximal-charge boundary vertex operator
\begin{displaymath}
h_{ \overline{\imath}_1 \cdots \overline{\imath}_s}^{
j_{s+1} \cdots j_n} 
\eta^{ \overline{\imath}_1 } \cdots \eta^{ \overline{\imath}_s }
\theta_{j_{s+1}} \cdots \theta_{j_n}
\end{displaymath}

Of course, by including the vacua in the discussion,
we find that if at least one of $\Lambda^{top} N_{S \cap T/T}$,
$\Lambda^{top} N_{S \cap T/S}$ is trivial, then one could
still get a maximal-charge boundary vertex operator
induced by the holomorphic top form on the Calabi-Yau.

Recall in section~\ref{pardiff} we ran into an apparent
problem with Serre duality.  At the time, we had not
taken into account the Freed-Witten anomaly.
Let us take a moment to
work through the details.
First, if $T \subseteq S$, then $N_{S \cap T/T} = 0$,
so for one string orientation we were consistent in
section~\ref{pardiff} to ignore the Freed-Witten anomaly,
and so the boundary state analysis in 
section~\ref{pardiff} need not be redone.
At the same time, $N_{S \cap T/S}  = N_{T/S}$, so we see
in our present language that if $\Lambda^{top} N_{T/S}$ is nontrivial,
then we would naively run into problems with Serre duality,
as indeed we saw in section~\ref{pardiff}.  By taking into account the
Freed-Witten anomaly,
we are able to restore Serre duality
invariance of the open string spectrum. 
Thus, we have
solved the puzzle presented in section~\ref{pardiff}.

\subsection{Proposal for new selection rule}  \label{newselection}

In this section, we shall make a proposal for a new 
selection rule for BPS brane configurations.
Specifically, we propose that whenever the 
line bundle 
\begin{equation}   \label{anom1}
\Lambda^{top} N_{S \cap T/T} \otimes
\Lambda^{top} N_{S \cap T/S}
\end{equation}
is nontrivial, the corresponding brane configuration is
non-BPS, when working near large radius, and when the
B field vanishes identically.

This proposal is motivated by our earlier
anomaly computation, that told us when the Chan-Paton factors have no curvature
and $TX|_S$ splits holomorphically, the open string B model is
only well-defined when the line bundle~(\ref{anom1}) is trivializable.
Recall this is the open string analogue of the statement that the closed string
B model is only well-defined for Calabi-Yau target spaces.

Let us check this statement empirically.
Suppose $S = X$, which is taken to be a Calabi-Yau
threefold with holonomy precisely equal to
$SU(3)$ (so in particular $X$ is not
$T^6$ or $K3 \times T^2$).  Let $T$ be a curve in $X$,
other than an elliptic curve.
In this case, $\Lambda^{top} N_{S \cap T/T}$ is trivial,
but $\Lambda^{top} N_{S \cap T/S} = \Lambda^{top}
N_{T/S}$ is nontrivial, so according to our analysis in
section~\ref{twoanom}, the open string B model is not well-defined
in this case.  If this brane configuration is supersymmetric,
then we appear to have a problem.
Now, the difference in dimensions
between $S$ and $T$ is a multiple of four, so naively this
brane configuration appears\footnote{Precisely 
at the large radius limit point, this brane configuration
{\it is} BPS.  Our remarks involving curvature of the
ambient space are irrelevant at the limit point, as the
space has become infinitely large and curvature has spread infinitely
thin.  However, if one is interested in results merely near large radius,
not actually at large radius, then our curvature considerations become
important. } to be BPS.
However, the ambient Calabi-Yau breaks too much supersymmetry.
After all, in a type II compactification, the ambient Calabi-Yau
leaves one with only ${\cal N}=2$ supersymmetry in four dimensions,
which is broken to ${\cal N}=1$ by the first brane.
However, ${\cal N}=1$ has no BPS states, so a second non-coincident
brane cannot be a BPS configuration.\footnote{It is possible in principle
for $S$ and $T$ to preserve precisely the same supersymmetry hence be
mutually BPS, although this is clearly non-generic.  Our assertion is that
this can only happen if $T$ is an elliptic curve.}
Thus, these two branes for which the open string B model
is not well-defined, are also not mutually supersymmetric. 

Similarly, if $S = X$, a Calabi-Yau threefold as above,
and $T$ is a divisor in $X$, then $\Lambda^{top} N_{S \cap T/T}$
is trivial, but $\Lambda^{top} N_{S \cap T/S}
= K_T$
is nontrivial (unless $T$ is itself Calabi-Yau)
and again the brane configuration appears to
be generically non-BPS, although this time the reason is much
more basic, namely the difference in dimensions is not a multiple
of four.  Thus, we have another example where the open string B model
is not well-defined, and the corresponding brane configuration is
non-BPS, consistent with expectations.

In every example of which we are aware in which the brane
configuration is BPS, the line bundle $\Lambda^{top} N_{S \cap T/T} \otimes
\Lambda^{top} N_{S \cap T/S}$ is trivializable,
so that the open string B model is not anomalous.
For example, consider parallel coincident branes 
{\it i.e.}, $S = T$, then both $N_{S \cap T/S} = 0$ 
and $N_{S \cap T/T} = 0$.
Similarly, if $T$ is a point on $S = X = K3$, then again
both $\Lambda^{top} N_{S \cap T/T}$ and
$\Lambda^{top} N_{S \cap T/S}$ are trivial, 
consistent with the
fact that the corresponding branes are mutually supersymmetric.

Note also that it is possible to have non-BPS configurations
such that the line bundle $\Lambda^{top} N_{S \cap T/T} \otimes
\Lambda^{top} N_{S \cap T/S}$ is trivializable -- we are claiming
that this line bundle defines a sufficient but not necessary
condition for a brane configuration to be non-BPS.
For example, if $S = X$ and $T$ is a divisor that is
also itself a Calabi-Yau manifold, then $\Lambda^{top} N_{S \cap T/T} \otimes
\Lambda^{top} N_{S \cap T/S}$ is trivial, yet this is clearly a non-BPS
configuration for dimension reasons.  So we are 
not conjecturing that a brane configuration is supersymmetric
if and only if both of those line bundles are trivial.
Rather we are only making the weaker conjecture
that if $\Lambda^{top} N_{S \cap T/T} \otimes
\Lambda^{top} N_{S \cap T/S}$ is nontrivial,
then close to large radius, with zero B field,  
the brane configuration will not be BPS.

\section{Nonintersecting branes}   \label{nonint}

In this section, for completeness
we shall very briefly discuss boundary spectra describing
open strings between D-branes on two completely disjoint complex
submanifolds of a Calabi-Yau manifold $X$.
Let the two submanifolds be denoted $S_1$, $S_2$, say, 
with inclusion maps $i_1$, $i_2$, respectively. 

In this case, if the two complex submanifolds are completely disjoint,
then there are no massless open string states connecting them.
Happily, it is also easy to check that in such circumstances,
all the groups
\begin{displaymath}
\mbox{Ext}^n_X\left( i_{1 *} {\cal E}_1 , i_{2 *} {\cal E}_2 \right)
\end{displaymath}
must vanish.  We can check this by noting that if there are no points
on $X$ where at least one of $i_{1 *} {\cal E}_1$, $i_{2 *} {\cal E}_2$
are zero, then all the corresponding local $\underline{\mbox{Ext}}$ sheaves  
must vanish, and so by the local-global spectral sequence,
the global Ext groups must all vanish as well.

\section{Ext groups of complexes}   \label{cpxes}

Another claim commonly made concerning the relationship
between D-branes and derived categories is that if open strings
with boundaries corresponding to two complexes should have
open string modes counted by Ext groups.
In other words, for an open string strip diagram,
if ${\cal E}_{\cdot}$ is a complex describing one boundary,
and ${\cal F}_{\cdot}$ is a complex describing the other boundary,
then open string modes should be counted by elements of
\begin{displaymath}
\mbox{Ext}^n_{D(X)} \left( {\cal E}_{\cdot}, {\cal F}_{\cdot} \right) \: = \:
H^n {\bf R}\mbox{Hom } \left( {\cal E}_{\cdot}, {\cal F}_{\cdot} \right)
\end{displaymath}

One can ask how these groups are realized physically,
just as earlier in this paper we asked how Ext groups between
coherent sheaves could be realized physically.
We saw how Ext groups between coherent sheaves 
are realized by vertex operators.  
What is the analogous procedure
for Ext groups of complexes?

We shall consider simple configurations involving
only branes, no antibranes.  We will find that
boundary vertex operators can be used
to determine countably many possible
Ext groups between complexes.  It is tempting to conjecture
that this ambiguity is closely related to possible
reinterpretations of this calculation in terms of brane/antibrane
configurations; however, we shall not say anything further here.

\begin{figure}
\centerline{\psfig{file=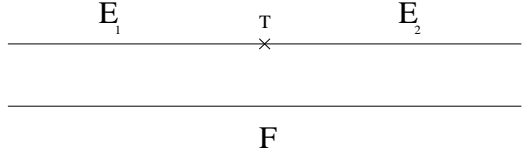,width=2.75in}}
\caption{\label{cpx1}Open string realizing map between simple complexes.}
\end{figure}

Let us consider the simplest possible nontrivial case,
in which 
\begin{displaymath}
{\cal E}_{\cdot}: \: \cdots \: \longrightarrow \: 0 \:
\longrightarrow \:
{\cal E}_1 \: \stackrel{T}{\longrightarrow} \: {\cal E}_2 \:
\longrightarrow \: 0 \: \longrightarrow \cdots
\end{displaymath}
and
\begin{displaymath}
{\cal F}_{\cdot}: \: \cdots \: \longrightarrow \: 0 \:
\longrightarrow \: {\cal F}_1 \: \longrightarrow \: 0 \:
\longrightarrow \: \cdots
\end{displaymath}
so ${\cal E}_{\cdot}$ has only two nonzero elements,
and ${\cal F}_{\cdot}$ has only a single nonzero element.
The corresponding open string diagram is shown in figure~(\ref{cpx1}).

Now, how can we see the states counted by $\mbox{Ext}( {\cal E}_{\cdot},
{\cal F}_{\cdot})$?  We shall loosely follow
the analysis of \cite{cardy2}.
The only boundary degrees of freedom are the asymptotic incoming
and asymptotic outgoing states.
A state coming in asymptotically from the left of
figure~(\ref{cpx1}) would only see boundaries ${\cal E}_1$
and ${\cal F}$; both ${\cal E}_2$ and the boundary operator $T$
would be effectively invisible.  Hence, assuming for simplicity
that ${\cal E}_1$ and ${\cal F}$ are both bundles on the same
submanifold $S$ of the Calabi-Yau $X$, our earlier analysis tells
us that the asymptotic incoming states are naively counted by the sheaf
cohomology groups
\begin{displaymath}
H^n \left(S, {\cal E}_1^{\vee} \otimes {\cal F} \otimes
\Lambda^m N_{S/X} \right)
\end{displaymath}
which determine (via a spectral sequence) elements of
\begin{displaymath}
\mbox{Ext}^{n+m}_X \left( i_* {\cal E}_1, i_* {\cal F} \right).
\end{displaymath}
Similarly, the asymptotic outgoing states (on the far right)
only see ${\cal E}_2$ and ${\cal F}$, and so are naively counted by
the sheaf cohomology groups
\begin{displaymath}
H^n \left(S, {\cal E}_2^{\vee} \otimes {\cal F} \otimes
\Lambda^m N_{S/X} \right)
\end{displaymath} 
which determine a corresponding Ext group.

Now, these asymptotic states determine an element of
the desired group
\begin{displaymath}
\mbox{Ext}^n_{ D(X) } \left( {\cal E}_{\cdot}, {\cal F}_{\cdot}
\right)
\end{displaymath}
as follows.
First, note that there is a short exact sequence of complexes
\begin{equation}   \label{secpx}
0 \: \longrightarrow \: {\cal E}_2 \: \longrightarrow \:
{\cal E}_{\cdot} \: \longrightarrow \:
{\cal E}_1[1] \: \longrightarrow \: 0
\end{equation}
an immediate consequence of the following trivial commuting diagram:
\begin{displaymath}
\xymatrix{
0: & \cdots \ar[r] & 0 \ar[r] \ar[d] & 0 \ar[r] \ar[d] & 0 \ar[r] 
\ar[d] & 0 \ar[d] \ar[r] & \cdots \\
{\cal E}_2: & \cdots \ar[r] & 0 \ar[r] \ar[d] & 0 \ar[r] \ar[d] & {\cal E}_2 \ar[r] \ar@{=}[d] &
0 \ar[d] \ar[r] & \cdots \\
{\cal E}_{\cdot}: & \cdots \ar[r] & 0 \ar[r] \ar[d] & {\cal E}_1 \ar[r]^{T} \ar@{=}[d] &
{\cal E}_2 \ar[r] \ar[d] & 0 \ar[d] \ar[r] & \cdots \\
{\cal E}_1[1]: & \cdots \ar[r] & 0 \ar[r] \ar[d] & {\cal E}_1 \ar[r] \ar[d] & 0 \ar[r] \ar[d] &
0 \ar[d] \ar[r] & \cdots \\
0: & \cdots \ar[r] & 0 \ar[r] & 0 \ar[r] & 0 \ar[r] & 0 \ar[r] & \cdots
}
\end{displaymath}
As a result of the short exact sequence~(\ref{secpx}),
we have a long exact sequence of Ext groups given by
\begin{eqnarray*}
\cdots & \longrightarrow & 
\mbox{Ext}^n_{ D(X) }\left( {\cal E}_1[1], {\cal F}_{\cdot} \right) \: 
\longrightarrow \:
\mbox{Ext}^n_{ D(X) }\left( {\cal E}_{\cdot}, {\cal F}_{\cdot} \right)
\: \longrightarrow \:
\mbox{Ext}^n_{ D(X) } \left( {\cal E}_2, {\cal F}_{\cdot} \right)
\: \longrightarrow \\
& \longrightarrow &  
\mbox{Ext}^{n+1}_{ D(X) } \left( {\cal E}_1[1], {\cal F}_{\cdot} \right)
\: \longrightarrow \: \cdots
\end{eqnarray*}
For the complex ${\cal F}_{\cdot}$ described above,
we can simplify these expressions.  The simplification depends
upon the relative grading of ${\cal E}_{\cdot}$ and ${\cal F}_{\cdot}$.
In the special case that ${\cal F}_1$ and ${\cal E}_2$ have the
same grading
\begin{eqnarray*}
\mbox{Ext}^n_{ D(X) } \left( {\cal E}_2, {\cal F}_{\cdot} \right)
& = &
\mbox{Ext}^n_X \left( {\cal E}_2, {\cal F}_1 \right) \\
\mbox{Ext}^n_{ D(X) } \left( {\cal E}_1[1], {\cal F}_{\cdot} \right)
& = &
\mbox{Ext}^{n-1}_X \left( {\cal E}_1, {\cal F}_1 \right)
\end{eqnarray*}
so we can rewrite the long exact sequence above more usefully
as follows:
\begin{eqnarray*}
\cdots & \longrightarrow &
\mbox{Ext}^{n-1}_X \left( {\cal E}_1, {\cal F}_1 \right) \:
\longrightarrow \:
\mbox{Ext}^n_{ D(X) } \left( {\cal E}_{\cdot}, {\cal F}_1 \right)
\: \longrightarrow \:
\mbox{Ext}^n_X \left( {\cal E}_2, {\cal F}_1 \right) \: \longrightarrow \\
& \longrightarrow & \mbox{Ext}^n_X \left( {\cal E}_1, {\cal F}_1 \right)
\: \longrightarrow \: \cdots
\end{eqnarray*}
More generally, if the grading of ${\cal F}_1$ is shifted $j$ units
to the left of ${\cal E}_2$, then
\begin{eqnarray*}
\mbox{Ext}^n_{ D(X) } \left( {\cal E}_2, {\cal F}_{\cdot} \right) 
& = &
\mbox{Ext}^{n+j}_X \left( {\cal E}_2, {\cal F}_1 \right) \\
\mbox{Ext}^n_{ D(X) } \left( {\cal E}_1[1], {\cal F}_{\cdot} \right)
& = &
\mbox{Ext}^{n-1+j}_X \left( {\cal E}_1, {\cal F}_1 \right)
\end{eqnarray*}
in which case we can rewrite the long exact sequence as
\begin{eqnarray*}
\cdots & \longrightarrow &
\mbox{Ext}^{n-1+j}_X\left( {\cal E}_1, {\cal F}_1 \right)
\: \longrightarrow \:
\mbox{Ext}^n_{ D(X) } \left( {\cal E}_{\cdot}, {\cal F}_{\cdot} \right)
\: \longrightarrow \:
\mbox{Ext}^{n+j}_X \left( {\cal E}_2, {\cal F}_1 \right) 
\: \longrightarrow \\
& \longrightarrow &
\mbox{Ext}^{n+j}_X \left( {\cal E}_1, {\cal F}_1 \right)
\: \longrightarrow \: \cdots
\end{eqnarray*}

Thus, we see that boundary vertex operators can be used
to determine Ext groups between complexes, but there is
an ambiguity in the grading.

\section{Conclusions}

In this paper we have explored recent claims that, for
D-branes wrapped on complex submanifolds of Calabi-Yau's,
open string states between D-branes are counted by Ext groups.
We have given much more detailed checks of this claim than
have appeared previously, and have worked out vertex operators
corresponding to Ext group elements in some generality.

In general terms, we have found that naively massless states in the
Ramond sector of open strings between intersecting D-branes
(wrapped on complex submanifolds, near large radius, with
zero B field) are in one-to-one correspondence with sheaf cohomology
groups, which are related to the desired Ext groups via spectral
sequences.  We have checked in a subclass of cases  
that those spectral sequences are realized physically via
BRST cohomology, ultimately because of
a Chan-Paton-induced modification of the open string boundary
conditions \cite{abooetal}.  We conjecture (though have
not been able to explicitly check) that the same is true
in general, that in all cases, the spectral sequences are
realized physically in BRST cohomology, so that
in general, massless Ramond sector states
are in one-to-one correspondence with Ext group elements.
These spectral sequences are nontrivial in general,
in the sense that the unsigned sum of the dimensions of the
sheaf cohomology groups is not the same as the unsigned sum of the
dimensions of the corresponding Ext groups, so understanding
their physical realization is an important issue.

For parallel (but not necessarily coincident) branes,
relating boundary vertex operators to Ext group elements is
straightforward physically.  However, for more general
brane intersections, we find a more interesting story. 
Specifically, we found that in order to be able to relate
boundary vertex operators to Ext groups,
we have to take into account the Freed-Witten anomaly,
which forces the gauge bundles on D-brane worldvolumes to
sometimes be twisted into non-honest bundles.
Not only does this allow us
to find a relationship with Ext groups, but it also fixes a
naive breakdown in Serre duality invariance of the spectrum.
Finally, we point out that a separate anomaly in the open string
B model, the open string analogue of the statement that closed
strings are only well-defined on Calabi-Yau's, yields 
a new (and very obscure) selection rule for BPS states.

In future work, we hope to return to the issue of the physical
realization of the spectral sequences in the remaining cases.
We conjecture that those spectral sequences are realized physically
via BRST cohomology, so that the massless Ramond sector states
are in one-to-one correspondence with Ext group elements,
but we have only checked this explicitly in the case of
parallel coincident branes.

\section{Acknowledgements}

We would like to thank P.~Aspinwall, R.~Bryant, S.~Dean,
F.~Denef, M.~Douglas,
J.~Evslin, D.~Freed, A.~Greenspoon, S.~Hellerman, A.~Knutson,
A.~Lawrence, D.~Morrison, R.~Plesser, E.~Scheidegger,
J.~Stasheff, R.~Thomas, C.~Vafa,
and E.~Witten for useful
conversations.

\appendix

\section{Derivation of spectral sequences}  \label{specseqap}

In this appendix, we give rigorous derivations of the
spectral sequences that are used in the paper.

\subsection{Parallel coincident branes}

Let $X$ be a complex manifold.  In our applications, $X$ will be Calabi-Yau
but this is not necessary so we do this more general situation which could
conceivably be of interest for more general topological string theories
than have been considered here.

Let $S$ be a smooth complex submanifold of $X$, and 
let $i:S\hookrightarrow X$ be the inclusion.  Finally, let
${\cal E}$ and ${\cal F}$ be bundles on $S$. The goal of this section is to
compute $\ext{p+q}_X(i_*{\cal E},i_*{\cal F})$, verifying the 
spectral sequence (\ref{easyss}) which we reproduce here for convenience:
\begin{displaymath}
E_2^{p,q}:
H^p \left(S, {\cal E}^{\vee} \otimes {\cal F} \otimes \Lambda^q N_{S/X} 
\right) \: \Longrightarrow \:
\mbox{Ext}^{p+q}_X\left( i_* {\cal E}, i_* {\cal F} \right)
\end{displaymath}

The method is to compute the local Ext sheaves $\uext^{*}(i_*{\cal E},
i_*{\cal F})$ (which
are supported on $S$), and
then use the local to global spectral sequence
\begin{equation}
\label{easylocglob}
H^p(X,\uext^{q}(i_*{\cal E},i_*{\cal F}))
\Longrightarrow \ext{p+q}_X(i_*{\cal E},i_*{\cal F}).
\end{equation}
Note that $H^p(X,\uext^{q}(i_*{\cal E},i_*{\cal F}))=H^p(S,
\uext^{q}(i_*{\cal E},i_*{\cal F}))$ when $\uext^{q}(i_*{\cal E},
i_*{\cal F})$ is viewed as a
sheaf on $S$.

Since
\begin{equation}
\label{easyreduce}
\uext^q(i_*{\cal E},i_*{\cal F})=\uext^q(i_*{\cal O}_S,i_*{\cal O}_S)\otimes
{\cal E}^\vee\otimes {\cal F},
\end{equation}
we can and will assume temporarily that ${\cal E}$ and ${\cal F}$ 
are both ${\cal O}_S$.  
Since $S$ is smooth, it is a local complete intersection \cite[P.~20]{gh},
so we can work locally and
assume that $S$ is the zero locus of a regular section $s\in
H^0(E)$ where ${E}$ is a bundle on $X$.  We will
eliminate dependence on ${E}$ in the results by noting
${E}|_S\simeq N_{S/X}$, as will be verified shortly.

We have the Koszul resolution
\begin{equation}
\label{koszul}
0\to \cdots \to \wedge^2 {E}^*\to{E}^*\to{\cal O}_X \to
i_*{\cal O}_S\to 0
\end{equation}
where all maps except the last restriction map
are defined as contraction by $s$.  
Explicitly, the map $\wedge^q E^\vee\to \wedge^{q-1} E^\vee$ is given by
$$(\omega_1\wedge\ldots\wedge\omega_q)\mapsto
\sum (-1)^{j-1}\omega_j(s)\omega_1\wedge\ldots\wedge\omega_{j-1}
\wedge\omega_{j+1}\wedge\ldots\wedge\omega_q.$$
The Koszul complex (\ref{koszul}) is exact if and only if $s$ is a regular
section, i.e.\ if and only if the rank of $E$ is equal to the 
codimension of $S$ in $X$.  See \cite{gh,eisenbud}.

For
later use, note that (\ref{koszul}) omitting ${\cal O}_X$ is
self-dual up to a twist: if ${E}$ has rank $r$ so that
$\wedge^r{E}$ is a line bundle, then
$\wedge^k{E}\simeq\wedge^{r-k}{E}^*
\otimes(\wedge^r{E})$.  See \cite[Proposition 17.15]{eisenbud}.

We can truncate (\ref{koszul}) to obtain the surjection
\begin{equation}
\label{trunc}
{E}^*\to{\cal I}_{S/X}
\to 0
\end{equation}
where ${\cal I}_{S/X}$ denotes the ideal sheaf of $S$ in $X$.  Tensoring
(\ref{trunc}) with $i_*{\cal O}_S={\cal O}_X/({\cal I}_{S/X})$ we
get the surjection ${E}^*|_S\to {\cal I}_{S/X}/({\cal I}_{S/X})^2$
which can be seen to be an isomorphism by using local equations for $S$ in 
$X$.  Since ${\cal I}_{S/X}/({\cal I}_{S/X})^2\simeq N_{S/X}^*$, we
see that ${E}|_S\simeq N_{S/X}$ as claimed.

We use (\ref{easyreduce}) and (\ref{koszul})  to calculate 
$\uext^*_{{\cal O}_X}(i_*{\cal E},i_*{\cal F})$ as the cohomology
sheaves of the complex
\begin{equation}
\label{localext}
\left(\wedge^*{E}\otimes{\cal E}^\vee\otimes
{\cal F}\right)|_S.
\end{equation}

Note that $s|_S=0$ by construction, so all maps in (\ref{localext}) are 0.
Combining with $E|_S\simeq N_{S/X}$, we get
\begin{equation}
\label{easyloc}
\uext^q_{{\cal O}_X}(i_*{\cal E},i_*{\cal F})\simeq
{\cal E}^{\vee} \otimes {\cal F} \otimes \Lambda^q N_{S/X}.
\end{equation}.

Then the claimed spectral sequence (\ref{easyss}) comes from substituting
(\ref{easyloc}) into the local to global spectral sequence (\ref{easylocglob}).

This spectral sequence has previously 
appeared in the string theory literature, e.g. \cite{paul}.

\subsection{Parallel branes of different dimension}

In this section we shall derive the spectral sequence
(\ref{intss}) which we reproduce here for convenience:
\begin{displaymath}
E_2^{p,q} \: = \: H^p \left(T, {\cal E}^{\vee}|_T \otimes {\cal F}
\otimes \Lambda^q N_{S/X}|_T \right) 
\: \Longrightarrow \:
\mbox{Ext}^{p+q}_X\left( i_* {\cal E}, j_* {\cal F} \right).
\end{displaymath}
Here $T$ is a complex submanifold of $S$,
which is a complex submanifold of $X$,
${\cal E}$ is a holomorphic bundle on ${\cal S}$,
${\cal F}$ is a holomorphic bundle on ${\cal T}$,
and $i: S \hookrightarrow X$, $j: T \hookrightarrow X$
are inclusions.

Now, recall that we can relate local $\underline{\mbox{Ext}}$ sheaves
to global Ext groups by the local to global spectral
sequence generalizing (\ref{easylocglob})
\begin{equation}
\label{locglob}
E_2^{p,q} \: = \: H^p\left(S, \underline{\mbox{Ext}}^q_{{\cal O}_S}
( {\cal S}_1, {\cal S}_2 ) \right)  \Longrightarrow 
\mbox{Ext}^{p+q}_S\left( {\cal S}_1, {\cal S}_2 \right) 
\end{equation}
which is valid for any coherent sheaves ${\cal S}_1, {\cal S}_2$ on $S$.

To derive the result formally, we shall first 
show how to compute local $\underline{\mbox{Ext}}$ sheaves
in terms of analogous data, then apply the local-global spectral
sequence (\ref{locglob}).

As in the previous section, we can work locally and assume that 
$S$ is the zero locus of a regular section $s$ of a bundle $E$ on $X$. Then
we have the Koszul resolution of ${\cal O}_S$:
\begin{displaymath}
\cdots \: \longrightarrow \:
\Lambda^2 E^{\vee} \: \longrightarrow \: E^{\vee} \: \longrightarrow \:
{\cal O}_X \: \longrightarrow \: i_*{\cal O}_S\to 0,
\end{displaymath}
where $E$ is a holomorphic bundle on $X$ of rank equal to the complex
codimension of $S$ in $X$, with a section $s$ whose zero set is $S$.
The bundle $E$ also has the property that $E|_S = N_{S/X}$.

To compute the sheaf $\underline{\mbox{Ext}}^q_{{\cal O}_X }
( i_*{\cal O}_S, j_* {\cal F}
)$, we use the Koszul resolution above to provide a projective
resolution of ${\cal O}_S$.  Thus, the local $\underline{\mbox{Ext}}$
sheaf desired is the degree $q$ cohomology sheaf of the complex
\begin{displaymath}
\underline{\mbox{Hom}}_{ {\cal O}_X }\left( {\cal O}_X, j_* {\cal F} \right) \:
\longrightarrow \:
\underline{\mbox{Hom}}_{ {\cal O}_X }\left( E^{\vee}, j_* {\cal F} \right)
\: \longrightarrow \:
\underline{\mbox{Hom}}_{ {\cal O}_X }\left( \Lambda^2 E^{\vee}, j_* {\cal F} \right)
\: \longrightarrow \: \cdots  .
\end{displaymath}
Since $j_*{\cal F}$ is supported on $T\subset S$ and $s|_S=0$, again
we have that all maps are 0 and so
\begin{displaymath}
\underline{\mbox{Ext}}^q_{ {\cal O}_X } \left( i_*{\cal O}_S,
j_* {\cal F} \right) \: = \:
j_* \underline{\mbox{Hom}}_{ {\cal O}_T } \left(
\Lambda^q N^{\vee}_{S/X}|_T, {\cal F} \right)\simeq\Lambda^qN_{S/X}|_T\otimes
{\cal F}.
\end{displaymath}

Locally on $X$ we can form a bundle $\overline{ {\cal E}}$ such 
that $\overline{{\cal E}}|_S = {\cal E}$, and by tensoring the
projective resolution of ${\cal O}_S$ with $\overline{ {\cal E}}$
and repeating the analysis above we immediately get the result
\begin{displaymath}
\underline{ \mbox{Ext}}^q_{ {\cal O}_X } \left(
i_* {\cal E}, j_* {\cal F} \right) \: = \:
j_* \underline{\mbox{Hom}}_{ {\cal O}_T } \left(
{\cal E}|_T \otimes \Lambda^q N^{\vee}_{S/X}|_T, {\cal F} \right)
=j_*\left(
\left({\cal E}^\vee\otimes N_{S/X}\right)|_T\otimes{\cal F}\right).
\end{displaymath} 

Finally, using the result \cite[Lemma III.2.10]{hartshorne} that
\begin{displaymath}
H^* \left(X, j_* {\cal F} \right)  \: = \:
H^* \left(T, {\cal F} \right)
\end{displaymath}
we see that
\begin{displaymath}
H^p \left(X, \underline{\mbox{Ext}}^q \left( i_* {\cal E}, j_* {\cal F}
\right) \right) \: = \:
H^p\left(T, \left({\cal E}^{\vee}\otimes \Lambda^m N_{S/X}\right)|_T 
\otimes {\cal F}  \right)
\end{displaymath}
which together with the local-global spectral sequence tells us that
we have the desired level two spectral sequence
\begin{displaymath}
E_2^{p,q}:H^p\left(T,\left({\cal E}^{\vee}\otimes \Lambda^m N_{S/X}\right)|_T 
\otimes {\cal F}  \right)
\: \Longrightarrow \:
\mbox{Ext}^{p+q}_X\left( i_* {\cal E}, j_* {\cal F} \right).
\end{displaymath}

\subsection{General brane intersections}

Let's now turn to the general case.  We have to interpret
(\ref{localext}), which up to tensoring with bundles 
is the dual of a Koszul complex on a (not necessarily
regular) section $s$ of ${E}\otimes j_*{\cal O}_T
={E}|_T$.  Koszul complexes are exact over the locus where the
section is regular, so in particular (\ref{localext}) is exact on the
complement of $S\cap T$.  In other words, the cohomology sheaves of
(\ref{localext}) are supported on $S\cap T$ as was already clear
geometrically since these compute $\uext^*(i_*E,j_*F)$.

If we restrict (\ref{trunc}) to $T$ we again get a surjection
\begin{equation}
\label{restrunc}
{E}^*|_T\to {\cal I}_{S\cap T,T}\to 0
\end{equation}
but the restriction of (\ref{restrunc}) to $S\cap T$, i.e.\ 
${E}^*|_{S\cap T}\to N^*_{(S\cap T)/T}$, while certainly
a surjection, need not be an isomorphism.  Since ${E}|_S\simeq N_{S/X}$
this further restriction of (\ref{restrunc}) leads to a 
surjection $N_{S/X}^*|_{S\cap T}\to N^*_{(S\cap T),T}$.
Letting $N=(N_{S/X})|_{S\cap T}$ and $N'=N_{S\cap T/T}$, this can be
rewritten as a surjection $N^\vee\to (N')^\vee$.  Dualizing, we see that
$N'$ is a subbundle of $N$.  
\footnote{The inclusion $N'\subset N$ can also be seen directly from geometry.
Consider the natural composition $\psi:T(T)|_{S\cap T}\to T(X)|_{S
\cap T}\to N$ of the natural inclusion and quotient.  
The kernel of $\psi$ at $p\in S\cap T$ consists of $T_pT\cap T_pS$; but
this is $T_p(S\cap T)$ since $S\cap T$ is a submanifold.  So 
$T(T)|_{S\cap T}/\mathrm{ker}(\psi)\simeq N'$ and we have the claimed
inclusion $N'\subset N$.}

Denote the codimension of $S\cap T$ in $T$ by $k$.  Since considerations
are local we can and will assume that  
$s|_T$ is a section of a rank $k$ subbundle $ E'\subset {E}|_T$ 
whose restriction to $S\cap T$ is the subbundle $N'\subset N$.  Note that
$s|_T$ is immediately seen to be a regular
section of ${E}'$ since its zero locus $S\cap T$ has codimension 
$k$.

So we see that (\ref{localext}) is up to tensoring with bundles the dual of
a Koszul complex on a section of $E|_T$
which is regular as a section of the subbundle ${E}'$.  
Let $\tilde{N}=N/N'$ be the bundle on $S\cap T$ introduced in (\ref{bundleb}).
We claim that the $q^{\scriptstyle{{\rm th}}}$ cohomology of this Koszul
complex is $\Lambda^k(N')\otimes\Lambda^{q-k}\tilde{N}$, so that 
$\uext^q(i_*{\cal O}_S,
j_*{\cal O}_T)=\wedge^k(N')\otimes \Lambda^{q-k}(\tilde{N})$.  Thus
\begin{equation}
\label{final}
\uext^q(i_*{\cal E},
j_*{\cal F})=\Lambda^k(N')\otimes \Lambda^{q-k}(\tilde{N})\otimes({\cal E}
|_{S\cap T})^*
\otimes {\cal F}|_{S\cap T}.
\end{equation}
Then (\ref{final}) immediately leads to the spectral sequence claimed in
Section~\ref{correct} by considerations of vertex operators.

It remains to explain our claim.  This is justified by linear algebra and
local coordinates.  

Rather than give a careful proof, we content ourselves with explaining the
idea.  We can do this most easily 
if we assume that $E|_T$ splits holomorphically 
into a direct sum
$E'\oplus E''$ with $E''|_{S\cap T}\simeq \tilde{N}$.

The only cohomology of the Koszul complex $\Lambda^\bullet
(E')^\vee$ is on the far right giving ${\cal O}_{S\cap T}$.
So the only cohomology of the dual complex $\Lambda^\bullet E'$ is on the
far right; by the self-duality we have mentioned earlier, we use
$\Lambda^\bullet E'\simeq \Lambda^\bullet (E')^\vee\otimes \Lambda^k E'$
to compute the cohomology of the dual complex as $\Lambda^k E'|_{S\cap T}
=\Lambda^kN'$.

Now using the full bundle $E$ rather than $E'$, we
note that
\begin{equation}
\label{split}
\Lambda^qE|_T=\bigoplus_i\Lambda^iE'\otimes\Lambda^{q-i}E''.
\end{equation}

The dualized Koszul complex then decomposes into a direct sum of the dualized
Koszul complex on $E'$ tensored with various $\Lambda^*E''$.
Computing cohomology and using $E''|_{S\cap T}\simeq \tilde{N}$, we get 
$\Lambda^k(N')\otimes \Lambda^{q-k}(\tilde{N})$.  Then we tensor with
$({\cal E}
|_{S\cap T})^*
\otimes {\cal F}|_{S\cap T}$ to arrive at (\ref{final}) as claimed.

For the general case, we can 
show that $\Lambda^p{E}|_T$ has a natural filtration with
graded quotients
$\Lambda^i{E}'\otimes\Lambda^{p-i}({E}|_T/
{E}')$.  Its restriction to $S\cap T$ is again
$\Lambda^i(N')\otimes\Lambda^{p-i}(\tilde{N})$.
This filtration can be used to modify the argument that we gave above.

As an interesting aside,
note that this spectral sequence is closely
related to a standard adjunction calculation in algebraic geometry.
For any complex manifold $Y$, if $Z$ is a complex submanifold
of complex codimension $r$, then it is straightforward to
show \cite[section 5.3]{gh} that
\begin{equation}
\label{dualize}
\underline{\mbox{Ext}}^q_{{\cal O}_Y}
\left( {\cal O}_Z, K_Y \right) \: = \:
\left\{ \begin{array}{rl} 
        0 & q < r \\
        K_Z & q = r   \end{array}  \right.
\end{equation}

This also follows readily from our computations above.  By taking
determinants in the exact sequence 
$$0\to T(Z)\to T(Y)|_Z\to N_{Z/Y}\to 0$$
we see that $K_Z\simeq (K_Y)|_Z\otimes \Lambda^{\rm top}N_{Z/Y}$. 
We now can compare (\ref{dualize}) to (\ref{final}) with
$S=Z$, $T=X=Y$, ${\cal E}={\cal O}_Z$ and ${\cal F}=K_Y$.  Then
$S\cap T=Z$, $N'=N_{Z/Y}$, and $\tilde{N}=0$.  Then (\ref{final}) becomes
precisely (\ref{dualize}) for $q\le r$.

\end{document}